\documentclass[journal=jpcafh,manuscript=article]{achemso}

\usepackage[version=3]{mhchem} 
\usepackage{caption}
\usepackage{subcaption}
\usepackage{color, colortbl}
\usepackage{amssymb}
\usepackage{amsmath}

\newcommand*{\up}{\textsuperscript}

\newcommand*{\mean}[1]{\left< #1 \right>}
\newcommand*{\diff}[1]{\mathop{\mathrm d #1}}
\newcommand*{\kT}{k_{\rm B}T}

\newcommand{\SItrajdistmat}          	{S1} 
\newcommand{\SITrypsankey}       	{S2} 
\newcommand{\SItrueG}      	 	{S3} 
\newcommand{\SIpcaaas}       		{S4} 
\newcommand{\SIpcstr}        		{S5}
\newcommand{\SIeig}      			{S6} 
\newcommand{\SIpccomp}          	{S7}  
\newcommand{\SImlacurr}          	{S8} 
\newcommand{\SImlcontacts}          	{S9} 

\newcommand{\SImlimportance}       {S1} 
\newcommand{\SIH}       			{S2} 

\author{Simon Bray}
\altaffiliation{Bioinformatics Group, Institute of Informatics, University of Freiburg, Freiburg, Germany}
\author{Victor T\"anzel}
\affiliation[University of Freiburg]
{Biomolecular Dynamics, Institute of Physics, University of Freiburg, Freiburg, Germany}
\author{Steffen Wolf}
\email{steffen.wolf@physik.uni-freiburg.de}
\phone{+49 (0)761 203 5913}
\fax{+49 (0)761 203 5883}
\affiliation[University of Freiburg]
{Biomolecular Dynamics, Institute of Physics, University of Freiburg, Freiburg, Germany}

\title[ML and Networks]
  {Ligand unbinding pathway and mechanism analysis assisted by machine learning and graph methods}

\abbreviations{MD,dcTMD,ML,PCA,conPCA,UPGMA}

\begin{document}

%
%

\begin{abstract}
We present two methods to reveal protein-ligand unbinding mechanisms in biased unbinding simulations by clustering trajectories into ensembles representing unbinding paths. The first approach is based on a contact principal component analysis for reducing the dimensionality of the input data, followed by identification of unbinding paths and training a machine learning model for trajectory clustering. The second approach clusters trajectories according to their pairwise mean Euclidean distance employing the neighbor-net algorithm, which takes into account input data bias in the distances set and is superior to dendrogram construction. Finally, we describe a more complex case where the reaction coordinate relevant for path identification is a single intra-ligand hydrogen bond, highlighting the challenges involved in unbinding path reaction coordinate detection.
\end{abstract}

\section{Introduction}

Protein-ligand complex formation and dissociation\cite{Schuetz2017,bruce18,DeBenedetti2018} are a current focus in the fields of biomolecular simulations and non-equilibrium statistical mechanics, as understanding their microscopic details is of considerable pharmaceutical interest\cite{Swinney2012,copeland15}. Due to the time scales involved being several magnitudes outside of the capabilities of common all-atom molecular dynamics (MD) simulation methods, a range of biased simulation methods have been developed to enforce ligand unbinding such as infrequent metadynamics\cite{tiwary15,Shekhar2022}, random acceleration MD\cite{Luedemann2000,kokh18}, scaled MD\cite{Tsujishita1993,schuetz18,Bianciotto2021} or weighted ensemble MD\cite{Huber1996,Votapka2017}. In a similar vein, we have developed the dissipation-corrected targeted MD (dcTMD) method\cite{Wolf18,Post22}. Enforcing a constant unbinding velocity by a constraint force, dcTMD allows the calculation of potentials of mean force (PMF) $\Delta G$ and friction profiles $\Gamma$ for unbinding paths of protein-host complexes\cite{Wolf20,Jaeger22} from the resulting unbinding work $W$. Besides providing detail on the unbinding mechanism, these fields can further serve as input for the numerical integration of a Markovian Langevin equation\cite{Zwanzig01,Bussi2007} to reach simulation time scales within a biomedically relevant range of minutes and more.

The dcTMD method is based on a 2\up{nd}-order cumulant expansion of the Jarzynski identity\cite{Jarzynski97,Hendrix01} and thus requires a normally distributed $W$. While this approximation holds for isotropic cases such as liquids\cite{Post22}, we have found an imprudent application to biomolecular simulation data to result in a significant overestimation of friction and a subsequent underestimation of the potential of mean force\cite{Wolf20,Jaeger22}. Separating trajectories into ensembles that share similar characteristics in their dynamics during unbinding and calculating potentials of mean forces for each ensemble separately recovers a normally distributed $W$ for each ensemble and removes the overestimation artefact. Our experience is that the relevant characteristics involve distinct unbinding routes\cite{Wolf20} or protein conformational changes\cite{Jaeger22} that cause similar fluctuations experienced by the ligand and hence similar friction\cite{Wolf18}. Lumping together trajectories with different unbinding characteristics leads to a deviation from a normally distributed $W$ and cause the overestimation of friction\cite{Jaeger22}.

In this work, we describe two approaches that help a researcher to cluster trajectories from biased protein-ligand unbinding simulations into ensembles according to common characteristics and to reveal the underlying unbinding mechanisms in form of reaction coordinates, e.g., combinations of protein-ligand contact distances\cite{Ernst15} or protein backbone angles\cite{Jaeger22}. Projecting the simulation trajectories onto such reaction coordinates reveals distinct distributions in the respective histograms connecting start and end states. In the following, we define these connecting distributions as unbinding paths.

Identifying process paths in biomolecular simulations is a ubiquituous problem.\cite{Bolhuis02,Rohrdanz13,Lee17,Henin2022,Nguyen2022} In a more general sense, the problem is related to clustering trajectories of moving objects\cite{Yuan17}. Ligand unbinding paths can be determined e.g. by SEEKR\cite{Votapka17}, contact fingerprint analysis\cite{Lung2017,NunesAlves2021,Bianciotto2021}, volume-based metadynamics\cite{Capelli19}, adaptive bias potentials\cite{Rydzewski19} or, as in our case, via a principal component analysis of protein-ligand contacts (conPCA)\cite{Ernst15,Post19}.
Choosing the latter approach already presumes that protein-ligand contacts form the relevant coordinates do observe distinct paths, which may not necessarily be the case. Sorting trajectories by visual inspection can be tedious due to the necessity of manually inspecting hundreds of simulations\cite{Wolf20}. Furthermore, sorting can become ambiguous, if there is considerable overlap of trajectories from distinct pathways, which appears on shallow or rugged free energy surfaces. 
Commonly used dendrogram-based clustering approaches such as UPGMA\cite{Sokal1958} or neighbor joining\cite{Saitou1987} cannot display the resulting ambiguity in the data and may prove misleading when defining clusters\cite{huson10}.
 Additionally, the bias employed in targeted MD simulations can lead to an artificial pseudo-stationarity, leading to crossings between different paths\cite{Wolf20,Jaeger22} which cause mixing of fluctuation characteristics within a trajectory and rendering trajectory analysis via dcTMD useless. We note that in principle, such trajectories could be cut into parts and individual parts then be attributed to different ensembles. However, this approach causes a significant complication of data evaluation, and we therefore refrain from doing so.

\begin{figure}[htb]
	\centering
	\includegraphics[width=0.95\textwidth]{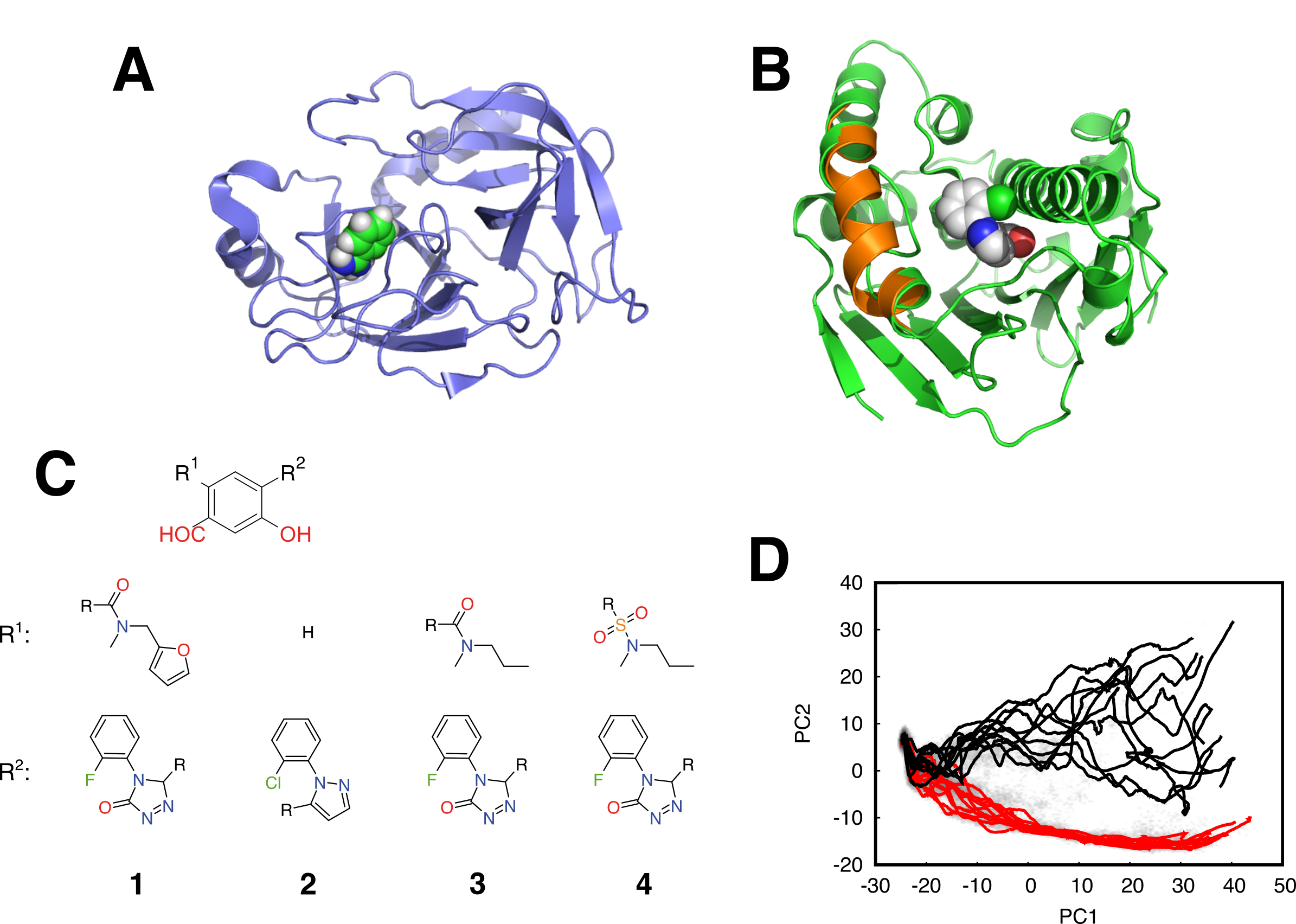}
	\caption{A: Trypsin-benzamidine complex. Protein as cartoon, benzamidine in spheres. B: Hsp90-ligand complex. Protein in ''loop binding'' conformation as green cartoon, in ''helix binding'' conformation in orange. Compound \textbf{2} in spheres. C: investigated Hsp90 compound set. D: Training data selected for classifying pathways of {\bf 1}, superimposed on a plot of the first two contact PCs. Path 1 in red, path 2 in black. Non-equilibrium biased energy landscape (see Eq.~(\ref{eq:NEQenergy})) as gray shading.} \label{fig:Intro}
\end{figure}

We demonstrate the different approaches on the well-established test systems of the Trypsin--benzamidine complex (Fig.~\ref{fig:Intro}a) and the N-terminal domain of heat shock protein 90 (Hsp90, Fig.~\ref{fig:Intro}b)\cite{schopf17,pearl06} with four sample compounds {\bf 1} to {\bf 4} (see Fig.~\ref{fig:Intro}b). For the second protein, all ligands share a common resorcinol functional group. If pathways can be readily observed for a subset of trajectories within a relevant coordinate space identified via conPCA, we provide a machine learning approach using the gradient boosting technique\cite{chen16,brandt18}. 
As an alternative to performing analysis based on pathways pre-defined by a human operator, we wanted to formulate a method that gives an unbiased estimate of similarities between trajectories, allowing for a bottom-up pathway identification similar to a density-based clustering\cite{sittel16} of MD data that is closer to the full dimensionality of the input data.
If initial pathways cannot be attributed, but a suitable distance metric between trajectories is known or can be guessed, we describe the usage of neighbor-nets\cite{Bryant2003,levy11} to cluster trajectories in order to identify possible paths. An advantage of using neighbor-nets is that it allows to take bias or ambiguities in the input data into account, which improves the quality of the trajectory assignment over clustering via dendrograms such as UPGMA\cite{sneath73,wheeler07}. Finally, we describe a case in which path separation has to be performed on the basis of a single ligand-internal hydrogen bond, highlighting the difficulty of finding good reaction coordinates for unbinding path analysis.

\section{Methods}

\subsection{Dissipation-corrected targeted molecular dynamics (dcTMD)}

To set the stage for the topic of this article, we briefly review the basics of dcTMD\cite{Wolf18}, the method for which pathway separation is required. Targeted MD makes use of a constant velocity constraint\cite{Schlitter94}
\begin{equation}
x(t) = x_0 + v_c t
\end{equation}
where $x$ is a position along a reaction coordinate of choice (in our case the distance between the centers of mass of two groups of atoms), $t$ is time and $v_c$ is the constraint velocity. 
The constraint is imposed by a constraint force $f_{c}$ calculated via a Lagrange multiplier for each time step. Integration of $f_{c}$ along $x$ results in a work $W(x)$ that is larger than $\Delta G (x)$.
According to Jarzynski's equality~\cite{Jarzynski97,Hendrix01}, $W$ and $\Delta{}G$ are related as
\begin{equation}
\Delta{}G = -\kT \ln{\langle e^{-W / \kT}\rangle} \approx \langle{}W\rangle - \frac{\langle{}\delta{}W^2\rangle}{2\kT}
= \langle{}W\rangle{} - W_{\rm diss},
\label{eq:Jarzy}
\end{equation}
where the mean $\mean{...}$ is calculated over an ensemble of trajectories initiated from an equilibrium distribution, and $W_{\rm diss}$ is the dissipative work. The approximation corresponds to a cumulant expansion truncated after the 2\up{nd} cumulant. Combining Eq.~(\ref{eq:Jarzy}) with a Markovian Langevin equation\cite{Zwanzig01}
allows the definition of a friction coefficient
\begin{equation}
\Gamma_{\rm NEQ}(x(t)) =  \frac{1}{k_{\rm B} T} \int_{0}^{t(x)} 
\left\langle \delta f_{\rm c}(t) \delta f_{\rm c}(t-\tau) \right\rangle 
\mathrm{d}\tau 
\label{eq:GammaNEQ}
\end{equation}
with $\delta f_{\rm c} = f_{\rm c} - \mean{f_{\rm c}}$ and $W_{\rm diss} = v_c \int_0^{x} \Gamma_{\rm NEQ}(x') \diff{x}$.

The major challenge in the applicability of dcTMD to a set of targeted MD trajectories lies in the validity of the approximation made when truncating the cumulant expansion, which only holds if $W$ follows a normal distribution\cite{Wolf18} within the set of trajectories. The presence of distinct pathways along an additional coordinate orthogonal to the bias coordinate leads to $W$ becoming multi-modal and to a significant overestimation of friction\cite{Wolf20,Jaeger22}. As this coordinate often is not known \textit{a priori} and must be identified for pathway separation, we denote it in the following as a ''hidden coordinate''. The typical method we use for revealing hidden coordinates is principal component analysis, which is introduced in the following.

\subsection{Principal component analysis (PCA)}
Principal component analysis (PCA) is a common method to reduce the dimensionality of a system to reveal relevant coordinates underlying a microscopical process of interest.~\cite{Ernst15,sittel18}. The method builds on the calculation of a covariance matrix 
\begin{equation}
\sigma{}_{mn} = \langle (r_m - \langle r_m \rangle) (r_n - \langle r_n \rangle) \rangle
\label{eq:PCA}
\end{equation}
from input coordinates $\mathbf{r}$. Diagonalizing this covariance matrix yields $i$ eigenvalues and -vectors $\mathbf{v^{(i)}}$ where $i$ is the number of unreduced dimensions of the system, describing the direction and variance of the principal motion. Usually, only the first few $\mathbf{v^{(i)}}$ with the largest eigenvalues that amount to $\gtrsim$80\% of cumulative eigenvalues
are chosen for further analysis. Projecting the original coordinates $\mathbf{r}$ onto the eigenvectors
\begin{equation}
x_i = \mathbf{r} \cdot \mathbf{v^{(i)}}
\end{equation}
yields the linearly uncorrelated principal components (PCs) $x_i$ that can serve as reaction coordinates. In the following, we use residue contact distances as $\mathbf{r}$. It should be noted that the definition of time-independent means $\mean{r_i}$ in Eq.~(\ref{eq:PCA}) in principle only applies to equilibrium MD data. Recently, PCA was extended to nonequilibrium TMD data\cite{Post2019}, and we used the $\mean{r_i}$ calculated over both time and ensemble of the nonequilibrium trajectories as reference. 

For contact analysis, we used all residues for which a minimum heavy atom distance of 4.5~\AA\ to the ligand (by center of mass) in any time frame in any pulling trajectory is reached. This results in e.g. 27 reference residues in Hsp90 that are displayed in Fig.~\SIpcaaas. These residue-ligand center-of-mass distances $\mathbf{r}(t)$ were subjected to PCA. Projections on the first three eigenvectors were plotted as histograms to reveal possible pathways in the form of ''valleys'' in their biased energy landscape\cite{Post2019,Lickert21} 
\begin{equation}
    \Delta \mathcal{G}_i = - \kT \ln (P(x_i)).
    \label{eq:NEQenergy}
\end{equation}
$\Delta \mathcal{G}$ is defined such that the underlying probability distributions can be qualitatively compared with potentials of mean force from equilibrium calculations.
An illustration of $\Delta \mathcal{G}$ and trajectories passing through two distinct pathways is provided by Figs.~\ref{fig:Intro}c and \SIpcstr a.

\subsection{Machine learning for pathway classification}

Following conPCA and visual identification of pathways in the PC space, we employed the XGBoost algorithm~\cite{chen16}, using the implementation developed by Brandt et al.\cite{brandt18}, to learn features of trajectories following individual pathways. The algorithm is supplied with preclassified training data from the ensemble, manually labeled according to pathway.
The selection of trajectories for training data was based on two factors: 1) trajectories should be unambiguously assignable by visual inspection to a single pathway, i.e. no crossover between pathways during the course of the trajectory; and 2) given the above, trajectories should nonetheless vary as much as possible. This ensures the model is trained with ''clean'' yet representative data.

A test/train split of 70/30 was employed. 100 training rounds were completed with learning rate set to 0.3 and maximum tree depth to 6 to construct a prediction model capable of assigning a pathway to unseen data. Training data was supplied as individual trajectory frames $\mathbf{r}(t)$ together with the visual pathway assignment. Data consisted of the 27 raw contacts used as predictor variables, and the manually assigned pathway as the target variable. As differences between the pathways were not clearly visible at the start of the trajectories, data in this initial region of the trajectories (in the case of Fig.~\ref{fig:Intro} c, all points with PC1 $<$ -15) were assigned to a ''neutral'' class which did not contribute to prediction, in order to avoid confusing the algorithm.

The model constructed was then used to predict the pathway taken by the remaining trajectories for all $\mathbf{r}(t)$. Once predictions were completed for individual time points, a score $S$ for each trajectory and path $j$ was calculated based on all $N$ time steps $i$ in a trajectory as
\begin{equation}
    S_j = \frac{1}{N} \sum_i^N \delta_j \quad \text{with} \, \delta_j =
    \begin{cases}
    1,& \text{if} \quad \mathbf{r}_i(t) \in j\\
    0,& \text{otherwise}.
    \end{cases}
\end{equation}
Points classified to the neutral state were ignored. All trajectories with $S_j > 0.8$ were classified as belonging to path $j$. Trajectories which did not meet this cut-off for any pathway remained unclassified.

\subsection{RMSD trajectory clustering \label{section:trajclust}}

As conPCA is based on contact distances between the ligand and residue center-of-masses, it is blind to changes in ligand conformation or rotation, which might also constitute hidden coordinates. Thus, efforts were made to develop a method capable of resolving such small-scale changes. To do so, we first aligned all trajectories based on a fit of the C$_\alpha$ atoms of the protein to provide a coordinate reference for ligand unbinding. After fitting, we calculated the time-dependent root mean square distance (RMSD) 
\begin{equation}
d_{ij}(t)=\sqrt{\frac{1}{N}\sum_{k=1}^{N}||\mathbf{l}_{jk}(t) - \mathbf{l}_{ik}(t)||^2}
\end{equation}
between any pair of trajectories $i$ and $j$ for all $N$ ligand atoms $k$ with Cartesian position vectors $\mathbf{l}$. 
Using raw values of $d_{ij}(t)$ is inadvisable, due to the increased motion of ligands after unbinding, as they randomly diffuse through the solvent. For large $t$, $d_{ij}(t)$ hence become large without providing information on the unbinding process. To mediate the influence of this drift, we generate normalized distances
\begin{equation}
\tilde{d}_{ij}(t)=\frac{ {d}_{ij}(t) }{ \mean{d(t)} }
\label{eq:RMSDs}
\end{equation}
Averaging all $\tilde{d}_{ij}(t)$ over time yields a RMSD matrix that encodes the dissimilarity of trajectories according to the relative position of ligand atoms, a representation of which is given for illustrative purposes in Fig.~\SItrajdistmat.

Having obtained the matrix, we used the distances to classify trajectories into clusters according to their similarity and interpreted these clusters as pathways. The problem is analogous to that of constructing a phylogenetic tree from genetic data depicting the evolutionary relationships between various species.~\cite{huson10} Two methods for clustering from distance data are therefore borrowed from the field of phylogenetics, which we introduce in the following.

\subsubsection{Unweighted pair group method using arithmetic averages (UPGMA)}
The unweighted pair group method using arithmetic averages (UPGMA) \cite{sneath73,wheeler07} is a bottom-up approach for construction of dendrograms, also known as phylogenetic trees, from a distance matrix~\cite{sneath73,wheeler07}. At first, each ''leaf'' of the tree (here, each trajectory) is considered to exist in its own cluster $C$ (called ''node'' in the following). The two nodes $i$ and $j$ which have the smallest $\tilde{d}_{ij}$ are merged into a single united node $k$. The distances between the new node $k$ and the remaining nodes are calculated as

\begin{equation}
\tilde{d}_{kl} = \frac{\tilde{d}_{il}|C_i| + \tilde{d}_{jl}|C_j|}{|C_i| + |C_j|}\quad \text{and} \quad |C_k| = |C_i| + |C_j|,
\end{equation}
where $|C_i|$ is the number of initial nodes, i.e., trajectories, in $i$, while $l$ is one of the other nodes that are not merged. The distance matrix is then updated with these new distances. Subsequently, node-merging is repeated iteratively on the respective two closest remaining nodes until all nodes are merged.

\subsubsection{Neighbor-net}

One major problem associated with dendrograms, whether calculated using UPGMA or a similar method such as neighbor-joining,\cite{Saitou1987} is that they do not take into account uncertainties in the input data. Especially if the elements in the distance matrix are similar to each other, inaccuracies caused, e.g., by problems with the initial superposition of the protein coordinates, might become dominant over the information in the individual distance pairs.\cite{huson10} In other words, dendrograms do not provide the equivalent of an error bar.

To allow such a representation, we here turn to the neighbor-net algorithm\cite{bryant04,levy11}. In this algorithm, the amalgamation procedure does not immediately unite a pair of nodes, but waits until the pair is united with a third node, upon which the amalgamation step replaces three nodes with two\cite{huson10}. 
Concretely, this entails selection of the two clusters (where a cluster may be either a single node or a node-pair) that minimize the adjusted distance $Q$, where
\begin{equation}
Q(C_i, C_j) = (m-2) \tilde{d}(C_i, C_j) - \sum_{k=1, k \neq{}i}^{m}\tilde{d}(C_i, C_k) - \sum_{k=1, k \neq{}j}^{m}\tilde{d}(C_j, C_k)
\end{equation}
Upon identification of the two clusters that minimize $Q$, the next step is to choose which node from each should be made neighbors (remembering each cluster may have either one or two members). The selection is based once again on which nodes $x_i$ and $x_j$ (where $x_i \in C_i$ and $x_j \in C_j$) minimize the adjusted distance $Q$, this time treating all members of $C_i$ and $C_j$ as if they were independent nodes:
\begin{equation}
Q(x_i, x_j) = (m-2) \tilde{d}(x_i, x_j) - \sum_{k=1, k \neq{}i}^{m} \tilde{d}(x_i, C_k) - \sum_{k=1, k \neq{}j}^{m} \tilde{d}(x_j, C_k)
\end{equation}
Once a node $y$ has two neighbors $x$ and $z$, it is replaced with two new nodes $u$ and $v$. The new distances between the new nodes and every other node $a$ are calculated by
\begin{align}
\tilde{d}(u, a) = [\tilde{d}(x,a) + \tilde{d}(y, a)] / 3 \\
\tilde{d}(v, a) = [\tilde{d}(y,a) + \tilde{d}(z, a)] / 3 \\
\tilde{d}(u, v) = [\tilde{d}(x,y) + \tilde{d}(x, z) + \tilde{d}(y, z)] / 3
\end{align} 
Iterating node merging until only two nodes remain (which are then linked in a final step) gives a circular splitting, as can be proven by induction~\cite{bryant04}.
In this way, a network is calculated, which forms a superposition of several possible dendrograms contained within the data.
If the input data exhibits a clear tree-like hierarchy, the neighbor-net collapses into a unique dendrogram. On the other hand, if ambiguity is present in the data, a network of equidistant nodes resembling a ''spider net'' appears, which represents the desired equivalent of an error bar. A decision can then be made where to impose cluster boundaries on the neighbor-net. 
An illustrative example is given in Ref.\cite{huson06}, Figs.~2 and 3: here, a dendrogram that apparently displays a clear and complicated node connection pattern collapses down to only two cluster of nodes in the neighbor-net, within which no further information on internode distance is available.

\subsection{Simulations of protein-ligand complexes\label{section:simdeets}}

Simulations were performed using the open-source Gromacs software~\cite{abraham15} v2018 using the AMBER99SB forcefield~\cite{Hornak06,Best06} and the TIP3P water model.~\cite{Jorgensen83} 
Simulation conditions for the trypsin-benzamidine complex are described in Ref.\cite{Wolf20} Here, we increased the number of pulling trajctories to a total of 400, each of 2~ns length at a pulling velocity of 1 m/s. Hsp90 simulations included four Hsp90-binding compounds named {\bf 1}, {\bf 2}, {\bf 3} and {\bf 4}, which are compounds {\bf 1b} (PDB ID 5J20)\cite{amaral17}, {\bf 1j} (PDB ID 6FCJ)\cite{gueldenhaupt18,Wolf19}, {\bf 1f} (modeled based on PDB ID 5J9X)\cite{amaral17} and {\bf 1g} (PDB ID 5J27)\cite{kokh18} in Ref.\cite{Wolf19}, respectively. The simulation systems and topologies were taken from Ref.\cite{Wolf19} Here, ligand topologies were created with antechamber\cite{wang06} and acpype\cite{Sousa2012} using GAFF parameters\cite{Wang04} and AM1-BCC charges\cite{Jakalian2000,Jakalian2002}. Simulations were performed using PME for electrostatics\cite{darden93} with a minimal real space cut-off of 1~nm and a van der Waals cut-off of 1~nm. Hydrogen atom bonds were constrained via the LINCS algorithm\cite{hess98}. For each ligand, statistically independent equilibration runs were performed (1000 for {\bf 1}, 500 for {\bf 2}, 100 for {\bf 3} and 513 for {\bf 4}, respectively) in the NPT ensemble at 300~K and 1~bar, using the Berendsen thermostat and barostat,~\cite{berendsen84} with an integrator time step of 2~fs and a trajectory length of 100~ps. Non-equilibrium TMD calculations using the Gromacs PULL code in constraint mode were then performed by continuing the equilibration runs for 2~ns in the NPT ensemble at 300~K and 1~bar, using the Nos\`e-Hoover thermostat~\cite{nose84} and Parrinello-Rahman barostat~\cite{parrinello81} with a fixed constraint velocity $v_c$ of 1~m/s and an integration step size of 1~fs. Values for the constraint forces were saved at each time step, while structural snapshots were written out each picosecond. The first pulling group was defined using all C$_{\alpha}$ atoms of the $\beta$-sheet forming the ligand binding site, while the second group was defined using the ligand heavy atoms (see Ref.\cite{Wolf19}, Fig.~S1).
For thermodynamic integration (TI)\cite{Berendsen07}, we saved $N=21$ structural snapshots at equidistant positions over the 2~nm pulling range and for each carried out 10~ns of constraint simulations with setting $v_c = 0$~m/s. For the last 5~ns, we calculated the time average of constraint force $\mean{f_c(x_i)}$ and from them a free energy profile 
\begin{equation}
\Delta G_{\rm TI}(x) \approx \sum_i^N \mean{f_c(x_i)} \Delta x
\label{eq:TI} 
\end{equation}
with spacing $\Delta x$ between the 21 snapshots.

\subsection{General data analysis}

Analysis was performed using the programming language Python, making particular use of the packages NumPy\cite{harris2020}, SciPy\cite{Virtanen2020} and Pandas\cite{mckinney10} for data analysis, as well as MDAnalysis\cite{agrawal11} for extracting data from MD files.
The XGBoost method was used in form of the xgbAnalysis package\cite{brandt18}. UPGMA trees were calculated using SciPy\cite{Virtanen2020}, while neighbor-nets are calculated and displayed via the implementation in SplitsTree\cite{huson06}.
Matplotlib\cite{hunter07} and Gnuplot\cite{gnuplot} were used for generating 2D plots, and Mayavi\cite{ramachandran11} for 3D plots. VMD\cite{humphrey96} and Pymol\cite{PyMOL15} were used for graphical visualization of structures and MD trajectories. Sankey plots were generated with PySankey (\url{https://github.com/vgalisson/pySankey}).

\section{Results and Discussion}

\subsection{Trypsin-benzamidine complex}

\subsubsection{Trajectory clustering based on RMSD using UPGMA}

\begin{figure}[htb]
\centering
\includegraphics[width=0.75\textwidth]{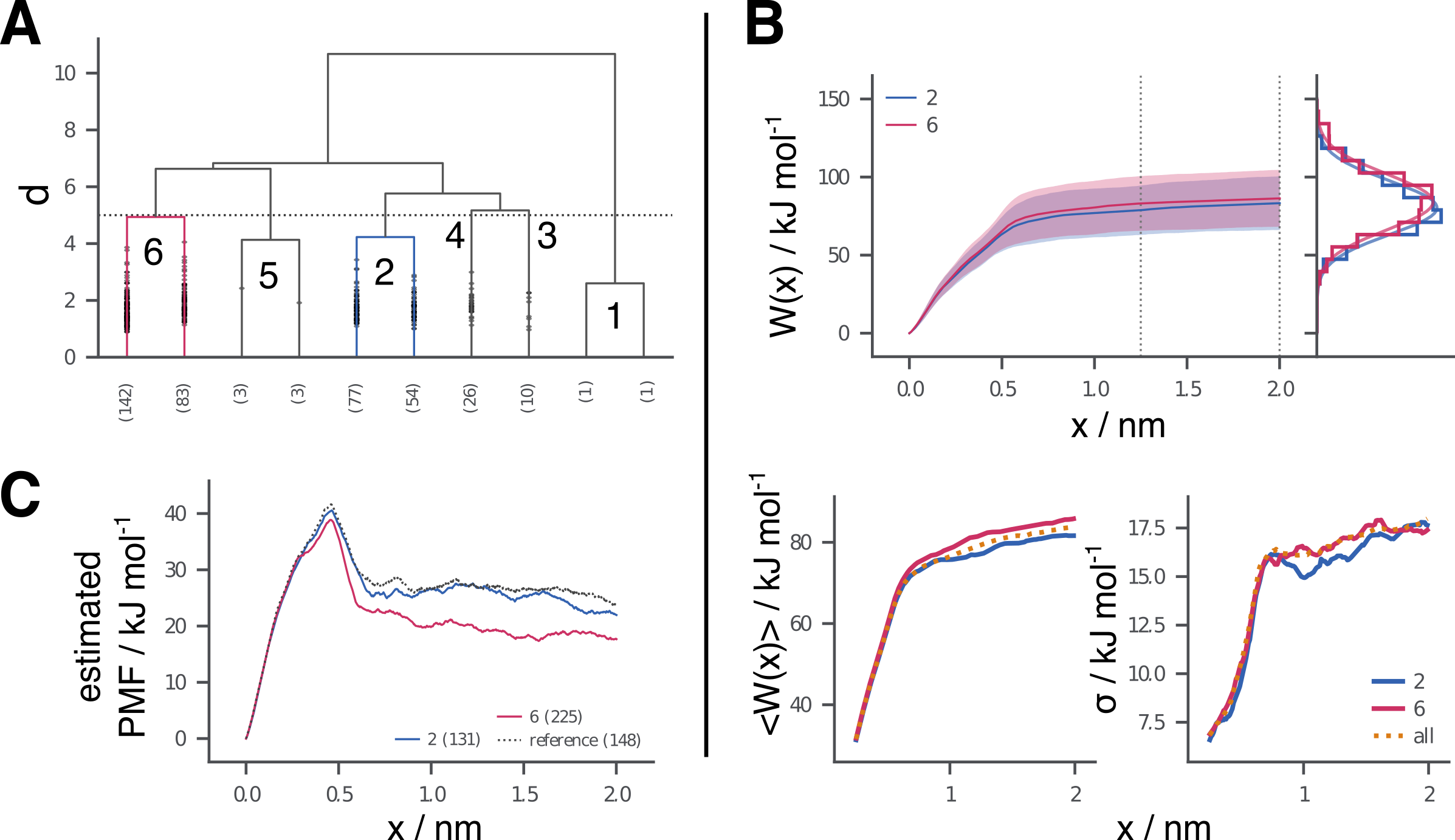}
\caption{UPGMA-based trajectory clustering for the trypsin-benzamidine complex. A: UPGMA dendrogram based on distance matrix according to Eq.~(\ref{eq:RMSDs}). Only the last splits for $d < 5$ are shown (additional splits are indicated by small grey circles). Number of trajectories contained in final nodes in brackets. B: Work distribution for trajectory clusters 2 and 6 together with mean values $\mean{W}$ as lines and standard deviation $\sigma$ as shades. Histograms show the distribution for $ x \in [ 1.25, 2.0 ] $~nm together with fits of normal distributions. C: Estimated potentials of mean force for pathways 2 and 6 together with a reference potential from Ref.\cite{Wolf20} }
\label{fig:TrypUPGMA}
\end{figure}

We begin this work with an unbinding path analysis of the well-understood trypsin-benzamidine complex\cite{tiwary15,Votapka17,Wolf20} to benchmark the distance matrix-based clustering approaches. Here, we have found paths to be contained within directions the ligand takes over the protein surface\cite{Wolf20} in agreement with other works\cite{tiwary15}. For all 400 simulations, a distance matrix was calculated according to Eq.~(\ref{eq:RMSDs}), and a UPGMA analysis performed, whose results are displayed in Fig.~\ref{fig:TrypUPGMA}. As the dendrogram in Fig.~\ref{fig:TrypUPGMA}a shows, setting a specific $d$ as cutoff value is highly arbitrary. We here decided for $d = 5$, yielding six clusters, of which only clusters 2 and 6 contain a sufficient number (more than 30, see Ref.\cite{Wolf18}) of trajectories for dcTMD analysis. Focusing on the distributions of $W$, Fig.~\ref{fig:TrypUPGMA}b shows that $W$ for both clusters of trajectories follow normal distributions that are heavily interleaved. Naively combining both clusters leads to a wrong estimate of $\mean{W(x)}$ and of the standard deviation $\sigma$, leading to an erroneous estimate of the potential of mean force in Eq.~(\ref{eq:Jarzy}). Comparison of the resulting PMF estimates in Fig.~\ref{fig:TrypUPGMA}c show that the estimate of cluster 2 is in good agreement with the PMF from our earlier works\cite{Wolf20}. This reference is based on a conPCA and visual inspection and gave good agreement with the experimentally determined binding rates and dissociation constant in Langevin simulation. However, a comparison between the trajectory clustering via UPGMA and via conPCA in Fig.~\SITrypsankey\ reveals that the UPGMA paths do not correspond to the paths identified in conPCA, where the ''middle'' pathway corresponds to the reference in Fig.~\ref{fig:TrypUPGMA}c, but display strong mixing. Obviously, trajectory clustering based on UPGMA can result in reasonably appearing estimates of potentials of mean force, but does not result in physically meaningful paths.

\subsubsection{Trajectory clustering based on RMSD using neighbor-net}

\begin{figure}[htb]
\centering
\includegraphics[width=0.75\textwidth]{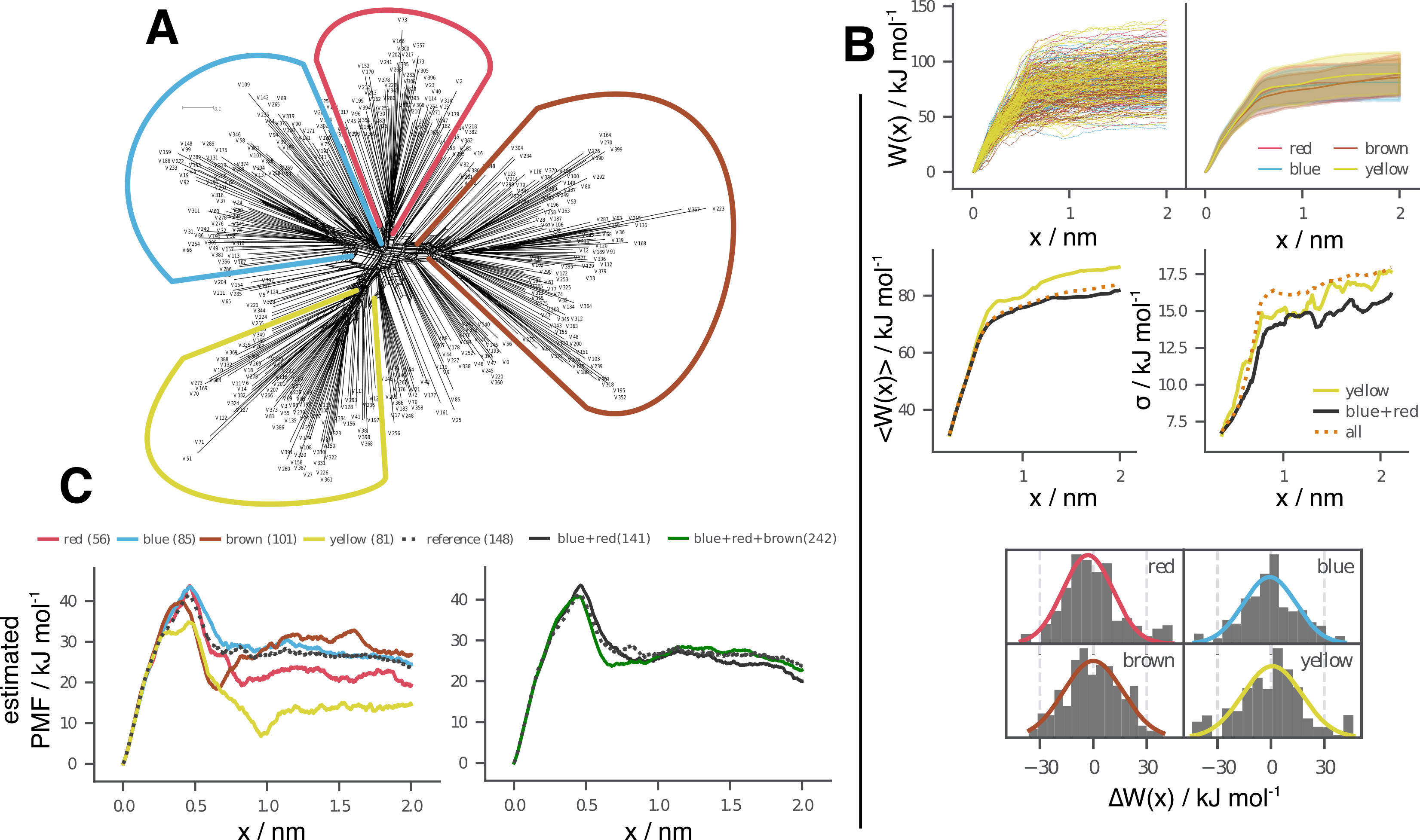}
\caption{Neighbor-net-based trajectory clustering for the trypsin-benzamidine complex. A: Neighbor-net based on distance matrix according to Eq.~(\ref{eq:RMSDs}) together with clusters defined. B: Work distribution for the six clusters. Shades display the standard deviation $\sigma$. Histograms show the distribution for $ x \in [ 1.25, 2.0 ] $~nm together with fits of normal distributions. C: Estimated potentials of mean force for pathways together with a reference value from Ref.\cite{Wolf20} } 
\label{fig:TrypNN}
\end{figure}

We now turn to neighbor-nets as clustering tool or RMSD data to see if the method results in physically meaningful ensembles of trajectories. Fig.~\ref{fig:TrypNN}a displays the resulting network and clusters attributed by us. In comparison to UPGMA, defining clusters still requires a visual analysis and choice, but choosing becomes simpler. Visually speaking, clusters are revealed as single ''branches'' that are separated by larger distances than the nodes connecting single trajectories, which form ''leafs'' sprouting from the branches. Fig.~\ref{fig:TrypNN}b displays the respective distributions of $W$, which are headily interleaved, and roughly follow normal distributions. While $\mean{W(x)}$ of the full trajectory set is in good agreement with the respective value for some of the identified ensembles, the variance estimate is too large, which causes an overestimation of friction. The resulting estimated potentials of mean force in Fig.~\ref{fig:TrypNN}c exhibit the best agreement with the reference profile\cite{Wolf20} for the ''blue'' cluster. Interestingly, lumping all clusters together except the ''yellow'' cluster retains good agreement with the reference. The comparison in Fig.~\SITrypsankey\ further clarifies that the blue cluster shows largest overlap with the reference ''middle'' pathway from conPCA and further includes some ''recrossing'' trajectories that could not be unambiguously attributed by visual inspection. On the other hand, the yellow cluster shows large overlap with the ''top'' cluster from conPCA. The neighbor-net clusters therefore can be related to the clusters from conPCA, giving them a physical meaning. Furthermore, the neighbor-net facilitates cluster definition in comparison to visually inspecting and comparing projections of single trajectories. 

In summary, it appears that neighbor-net outperforms UPGMA-based trajectory clustering and unbinding path definition. Similarly to the comparison between UPGMA and conPCA clusters, neighbor-net clusters do not exhibit clear overlap with UPGMA clusters, as well. One exception is the yellow cluster from neighbor-net, which is almost exclusively contained in cluster 6 from UPGMA. This observation coincides with yellow cluster and cluster 6 exhibiting notable deviations from the reference PMF (see Fig.~\SITrypsankey). It might be that at least for trypsin, our path separation results in good estimates of PMFs not by finding a single unbinding path, but by eliminating trajectories that go along an unphysical path. This is supported by the neighbor-net supercluster joining blue, red and brown clusters, yielding a reasonable PMF, as well. This supercluster contains trajectories from the bottom and middle clusters from conPCA as well as a large number of recrossing trajectories, but almost no trajectories from the top cluster.

\subsection{Hsp90-ligand complex}

\subsubsection{Pathway attribution for compound {\bf 1} by principal component analysis (PCA) and machine learning}

\begin{figure}[htb]
\centering
\includegraphics[width=0.75\textwidth]{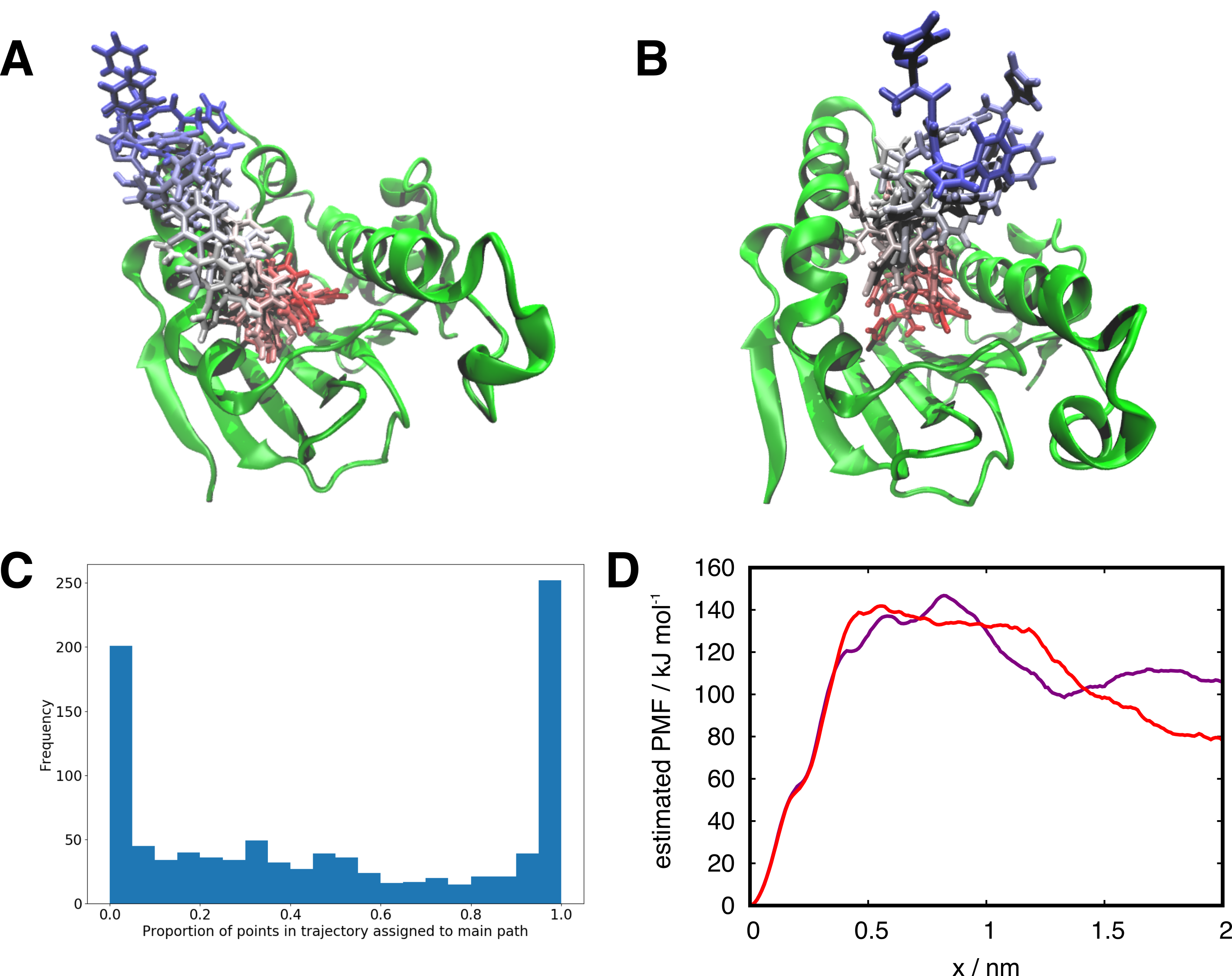}
\caption{A,B: Example trajectories for the two pathways 1 (A) and 2 (B). Ligand {\bf 1} positions are shown at 0.2~ns intervals and colored accordingly (red at the start of the trajectory, blue at the end.) C: Histogram of compound {\bf 1} ML scores for ensemble of 1000 trajectories for path 1. Values $>$ 0.8 correspond to path 1, values $<$ 0.2 to path 2. D: Estimated potentials of mean force for {\bf 1} pathways (from total ensemble of 1000 trajectories) as separated by PCA-ML (purple: path 1, 333 trajectories; red: path 2, 320 trajectories)} 
\label{fig:h42paths}
\end{figure}

For Hsp90, we initially developed and tested methods for pathway separation using compound {\bf 1} (see Ref.\cite{Wolf20}). 
As additional reference, we calculated a free energy profile via thermodynamic integration (TI) that is given in Fig.~\SItrueG. Here, we additionally observe that naively lumping all trajectories for dcTMD analysis results in an artificially low unbinding free energy of $\sim -100$~kJ/mol.
Figure~\SIeig\ shows cumulative eigenvalues for the principal components from conPCA, which reach 0.8 for the second and 0.95 for the third PC. As a result, two PCs (Fig.~\SIpccomp) are sufficient to describe the system, with additional detail provided by the third, which is easily rationalized as an internal coordinate representation of the three Cartesian coordinates the ligand can diffuse along.
Figures~\SIpcstr b and~\SIpcstr c show the location of the four residues that contribute most strongly to the value of the first two principal components. For PC1, these are located beneath the binding site: because it represents the largest variance in the data, PC1 correlates very strongly with the pulling coordinate. The residues that contribute most strongly to PC2 are concentrated at one edge of the binding site, so PC2 varies based on how close the ligand passes to this region.
Inspection of the coordinates by means of 2D and 3D plots revealed two main pathways for ligand dissociation (Fig.~\SIpcstr a and Supplementary Movie 1). The first two principal components suffice to distinguish the two pathways, though a three-dimensional plot including the third PC facilitates their identification. 
Pathway 1 passes close to the residues highlighted in Fig.~\SIpcstr c (low PC2, black line in Fig.~\ref{fig:Intro}c) and the pathway 2 far from it (high PC2, red line in Fig.~\ref{fig:Intro}c). Therefore, the two pathways, concretely considered, are two routes out of the protein on opposite sides of the binding site as displayed in Fig.~\ref{fig:h42paths}a,b.

We then used our machine learning procedure to score trajectories according to the pathway taken. Twenty trajectories (ten for each path; examples shown in Fig.~\ref{fig:Intro}d) were selected by visual inspection for model training, and the resulting model used to assign each point of every trajectory studied to one of the two pathways. Training a model using the XGBoost algorithm as described in Methods yielded an accuracy of 100\% on training data and 99.4\% on test data; as this is already enough to make high-quality predictions, we did not investigate other machine learning techniques further. A histogram of scores for path 1 is depicted in Fig.~\ref{fig:h42paths}c. As most of the values can be seen to cluster near 0 and 1, a sufficiently large class of trajectories could be obtained for both pathways, by selecting trajectories with $S_j > 0.8$ for path 1 and $S_j < 0.2$ for path 2 as described in the Methods, and a PMF was estimated using the dcTMD method.

Values for ''importance'' of each contact, i.e. the extent of its contribution to the model, can also be extracted\cite{brandt18}, permitting a comparison with the results from PCA. Fig.~\SImlacurr\ shows that three contacts (Phe134, Val136 and Val186, see Fig.~\SImlcontacts) in particular have high values (0.31, 0.29 and 0.26 respectively) summing to 0.86 (implying 86\% of the model's predictive power is derived from these three contacts, see Tab.~\SImlimportance). The importance of the remaining contacts drops sharply, with the fourth most important residue having a value of 0.03. Comparing with Fig.~\SIpccomp, residue 186 is one of the largest contributors to PC1, while contacts 134 and 136 are two of the largest contributors to PC2, showing the model has a similar opinion on the interpretation of the contact data as conPCA.

The estimated potentials of mean force for the PCA-ML pathways are given in Fig.~\ref{fig:h42paths}d. It can be observed that both pathways exhibit no friction overestimation artefact\cite{Jaeger22}, but differences in the estimated potential of mean force between bound and unbound states of $\sim$80--100~kJ/mol appear unrealistically large given an experimental\cite{amaral17} $K_{\rm D}$ of $\sim$5$\cdot$10$^{-9}$M\up{-1} (which, ignoring corrections\cite{Hall2020}, would correspond to a standard free binding energy $\Delta G_0 \sim$50~kJ/mol). 
Indeed, TI calculations with compound {\bf 1} (Fig.~\SItrueG) return an unbinding free energy of $\Delta G_{\rm TI} \sim$50--60~kJ/mol.
We therefore check in the following if RMSD-based clustering results in more reasonable potential of mean force estimate.

\subsubsection{Trajectory clustering from RMSD data}

\begin{figure}[htbp]
\includegraphics[width=0.75\textwidth]{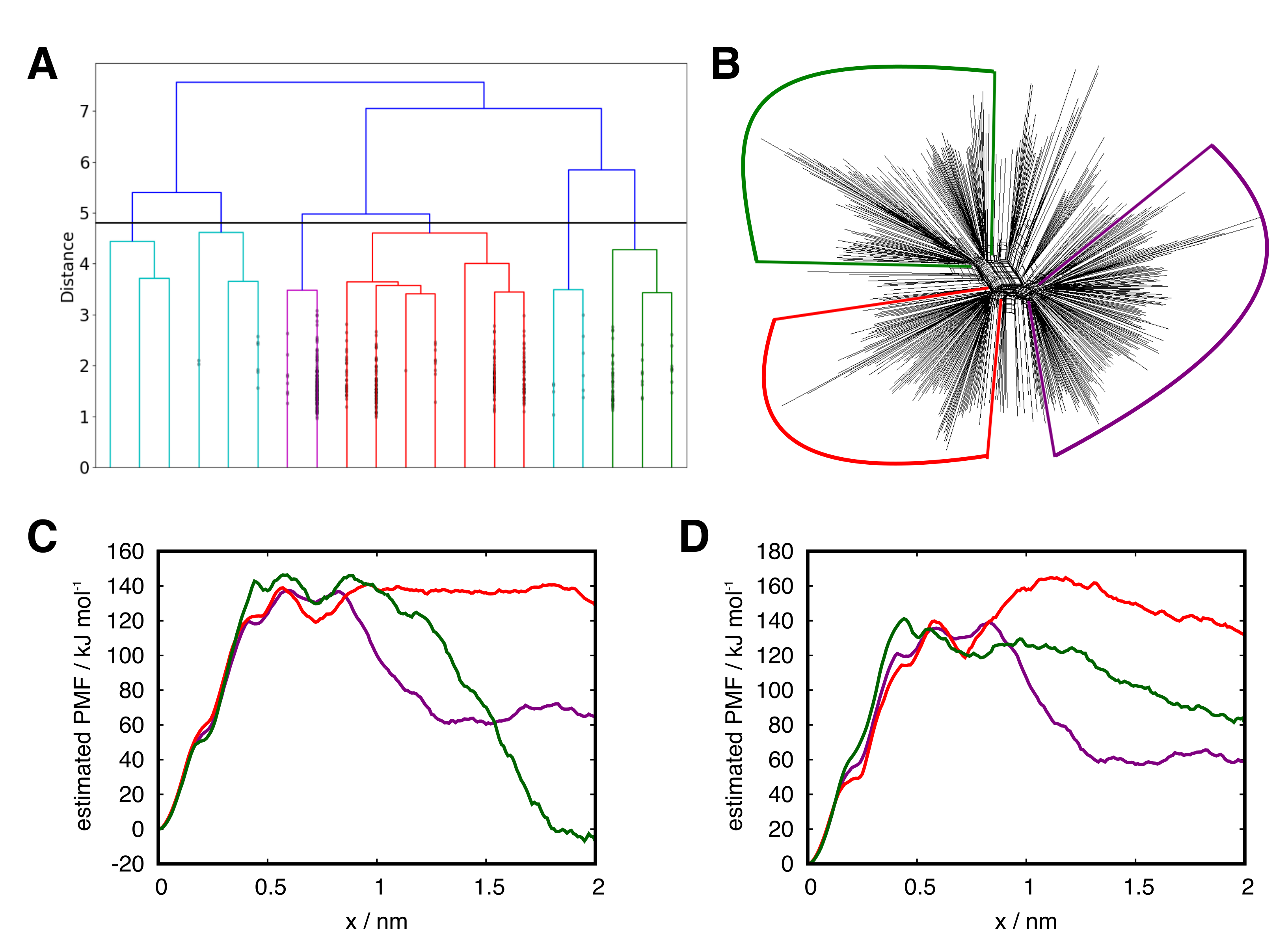}
\caption{A: Dendrogram of clustered {\bf 1} trajectories (only first 20 splits shown; additional splits are indicated by small grey circles). Imposing a cut-off at 4.8 (horizontal black line) gives six clusters, with populations of 177 (purple), 227 (red) and 62 (green); the remaining three, colored in cyan, have populations of 5, 14 and 14, and are not used for potential of mean force estimations. B: Network diagram produced by neighbor-net algorithm. The radiating lines represent nodes (i.e. trajectories), and the line lengths represent RMSD distances. Ambiguous node merging is visible in the center. Three main clusters are visible, while some trajectories (especially on the upper right) do not unambiguously fit into any larger cluster. Colored lines represent cluster boundaries chosen by a human operator. C: Estimated potentials of mean force for pathways from RMSD (\hbox{UPGMA}) clustering: path 1 (purple, population 177), path 2 (red, population 227), and path 3 (green, population 62)]. D: Estimated potentials of mean force for pathways from RMSD (neighbor-net) clustering: path 1 (purple, population 166), path 2 (red, population 106), and path 3 (green, population 124).}
\label{fig:H42_rmsd}
\end{figure}

For our RMSD-based clustering, $\tilde{d}_{ij}(t)$ distances were extracted from 500 trajectories of compound {\bf 1}, which yielded the distance matrix displayed in Fig.~\SItrajdistmat.
Hierarchical clustering using the UPGMA algorithm returned the dendrogram given in Fig.~\ref{fig:H42_rmsd}a. Based on the dendrogram, a cut-off distance of 4.8 was selected to yield several clusters, three of which are sufficiently populated ($>50$ trajectories)\cite{Wolf20} to carry out a potential of mean force estimation (Fig.~\ref{fig:H42_rmsd}c). 
In comparison to the PCA-ML approach, we observe the emergence of one additional pathway. Path 1 exhibits a final unbinding potential of mean force of $\sim$60~kJ/mol, which is more reasonable than the results from PCA-ML-based sorting. Path 2 exhibits a $\Delta G$ profile roughly similar to those found using PCA-ML, while the novel path 3 exhibits an unreasonable drop to negative values as in a friction overestimation artefact, which suggests at least some of the trajectories are still assigned wrongly.

\begin{figure}[htb]
\centering
\includegraphics[width=0.9\textwidth]{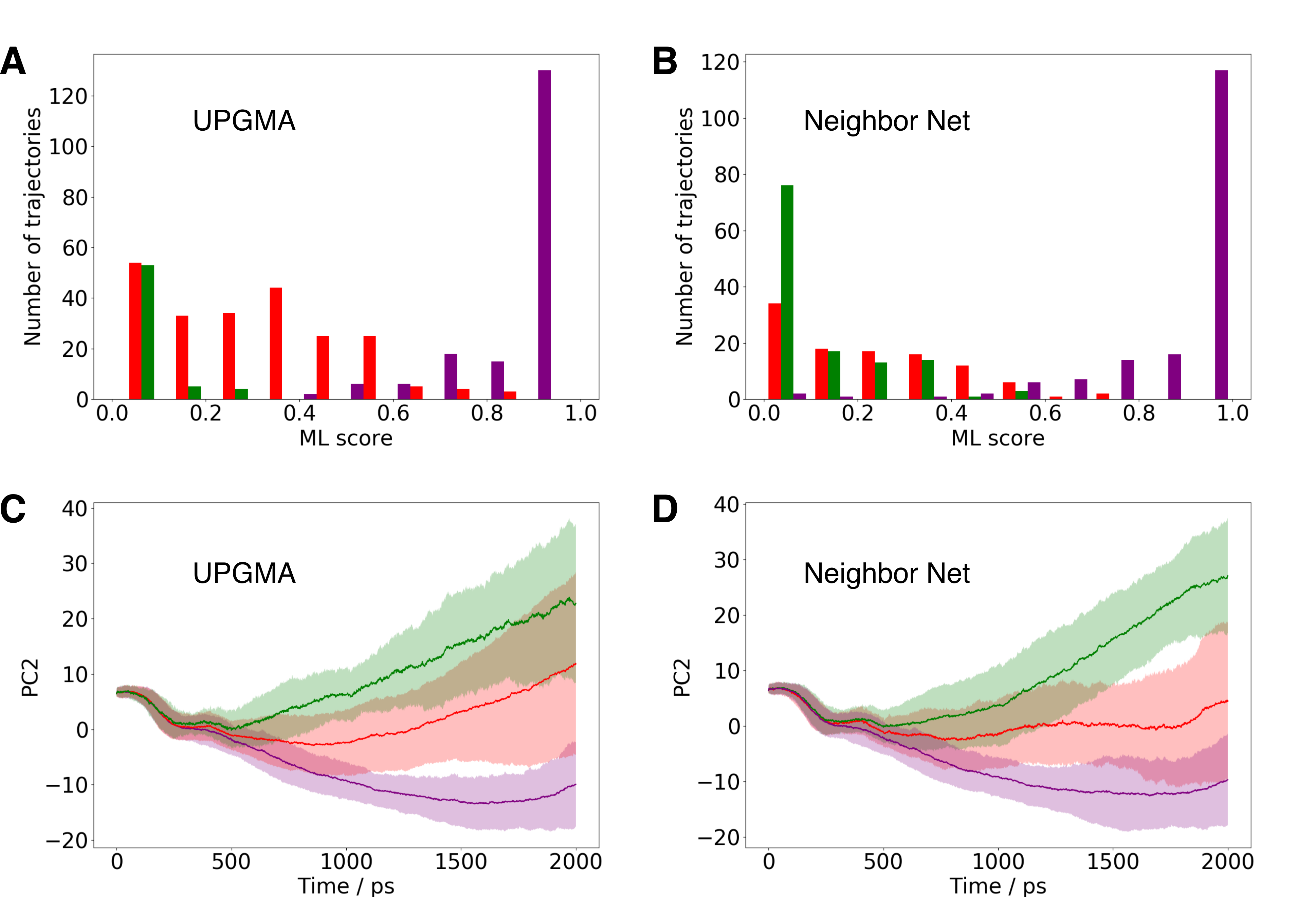}
\caption{A, B: ML scores for paths clustered by both \hbox{UPGMA} and neighbor-net as presented in Fig.~\ref{fig:H42_rmsd} using the same coloring scheme. C,D: Average PC2 value for RMSD clusters of trajectories plotted against time with standard deviation shading.} \label{fig:NeighborNet}
\end{figure}

To improve trajectory attribution, clustering using the neighbor-net algorithm was performed on the same data set as UPGMA clustering (500 trajectories), which yields the network in Fig.~\ref{fig:H42_rmsd}b. Three regions were identified where trajectories are more densely and unambiguously located, and on this basis three clusters are defined, excluding the remainder of the trajectories. Potentials of mean force were estimated for both classes, yielding results in Fig.~\ref{fig:H42_rmsd}d.
It can be seen that the path 1 (purple) dcTMD curve provides very similar information to path 1 produced by the UPGMA clustering. The two classes identified in each case for path 1 have almost identical composition (147 trajectories in common out of populations of 177 in UPGMA and 166 in neighbor-net). While path 2 (red) now exhibits an additional barrier at $x \sim$1.2 and ends with an unreasonably high final $\Delta G \sim$140~kJ/mol, the new path 3 (green) ends at a potential of mean force similar to path 1 and thus seems to be better resolved.

To check the source of the improved $\Delta G$ along path 3, we compare the pathways derived from RMSD clustering with the PCA-ML analysis. Figure~\ref{fig:NeighborNet}a,b shows the ML scores of the trajectories for each of the classes, demonstrating that path 1 (purple class) in both approaches corresponds to the PCA-ML path 1. 
Concerning paths 2 (red class) and 3 (green class), it turns out that path 3 actually corresponds to path 2 from PCA-ML, while path 2 represents a movement between both paths. 
Figs.~\ref{fig:NeighborNet}c,d indeed show that the ''green'' trajectories are those which follow the opposite side of the binding site to the PCA-ML path 1, while the ''red'' trajectories follow a more neutral course through the middle between both paths. This attribution explains the significantly higher $\Delta G$ of path 2, as the ligand is pushed through several protein residues, applying work to the protein fold. The neighbor-net algorithm indeed improves the attribution of trajectories towards path 2 and reduces the population of path-intermediate trajectories. 

Overall, RMSD clustering can be considered an improvement over the PCA-ML based data, with neighbor-net based clustering being superior to UPGMA clustering. 
A limitation of RMSD clusters however is the lack in information regarding the physical meaning of the pathways they represent. conPCA here is able to complement RMSD clustering, as it reveals more information on the Cartesian origins of such pathways.

\subsection{Pathway separation of further Hsp90 ligands}

\begin{figure}[htb]
\centering
\includegraphics[width=0.75\textwidth]{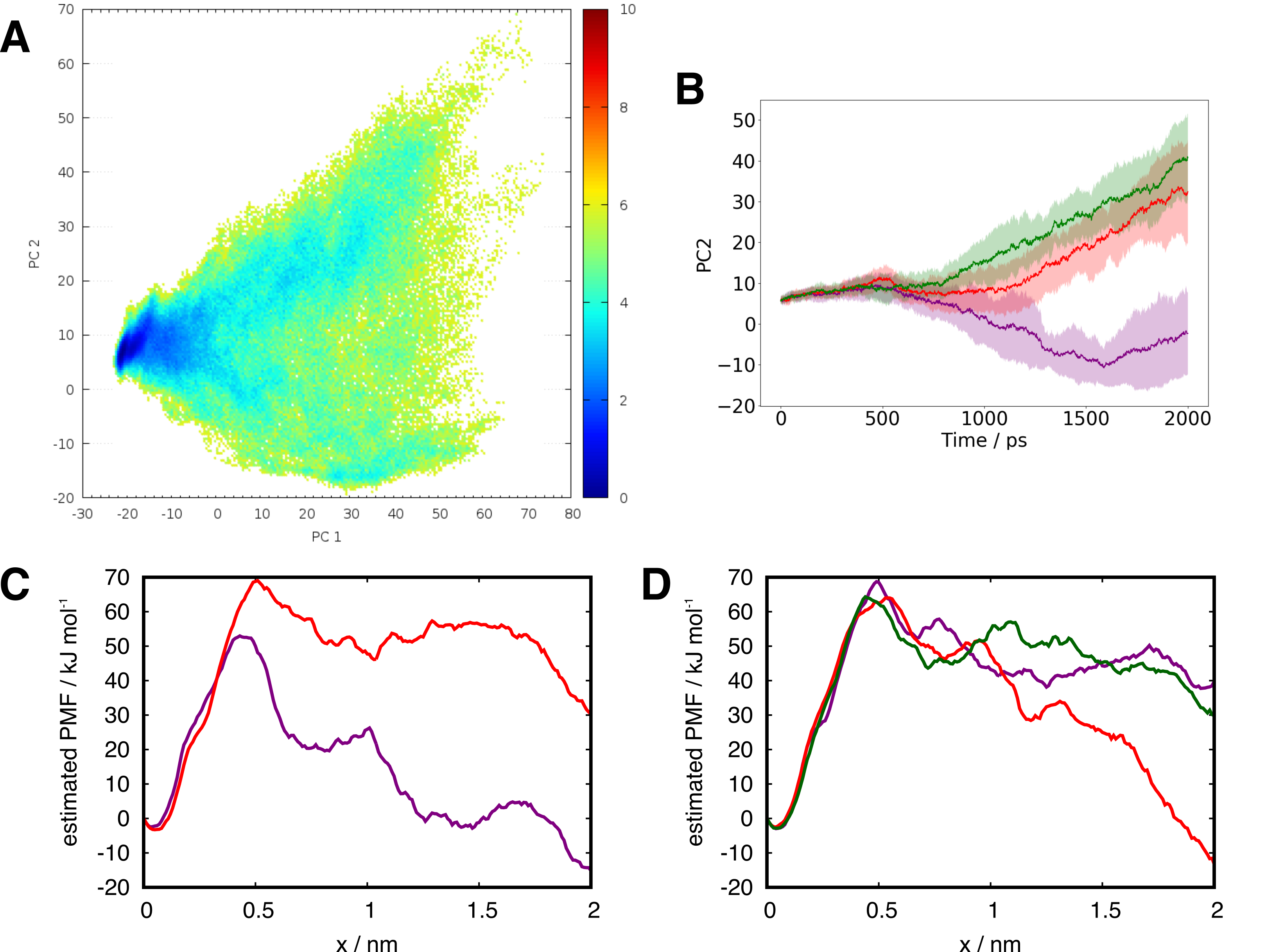}
\caption{A: Biased energy diagram\cite{Post2019} $\Delta \mathcal{G}$ for first two PCs of compound {\bf 2}. B: average PC2 value for RMSD/neighbor-net clusters plotted against time, with standard deviation as shading. Purple represents path 1, green and red represent two alternative definitions for path 2. C,D: estimated potential of mean force for (C) PCA-ML and (D) RMSD clusters.}
\label{fig:l22}
\end{figure}

Following testing of the pathway separation methods above with {\bf 1}, the applicability to the three other Hsp90 ligands {\bf 2}, {\bf 3} and {\bf 4} was tested. 
First of all, conPCA is performed, using the same contacts as for {\bf 1}, but diagonalizing the covariance matrix anew for each ligand, which provides different eigenvectors for each PCA. We focus first on compound {\bf 2}, which differs from {\bf 1} in that it binds to Hsp90 in the loop conformation, rather than the helix conformation. Here, the PCs again reveal two pathways. Unlike {\bf 1}, for which path 1 dominates, path 2 is far more heavily populated for {\bf 2}.

The PCA-ML pathway separation again yields two pathways, while RMSD/neighbor-net clustering splits one of the two paths into two sub-paths (see Fig.~\ref{fig:l22}a,b), as well.
As for {\bf 1}, the pathways pass along opposite sides of the binding site surface (Fig.~\ref{fig:l22}b). One of the two sub-paths (in red) gives unrealistic results, this time in the form of unreasonably low final $\Delta G$ values due to mixing of trajectories from different paths.
The results of {\bf 2} again suggest RMSD clustering provides better pathway separation than PCA-ML. Indeed, we used path 1 (purple in Fig.~\ref{fig:l22}d) identified via RMSD clustering in our Langevin equation simulations and analysis on binding and unbinding rate constants as well as $K_D$ and found good agreement with experimental values\cite{Wolf20}.

\begin{figure}[htb]
	\centering
	\includegraphics[width=0.75\textwidth]{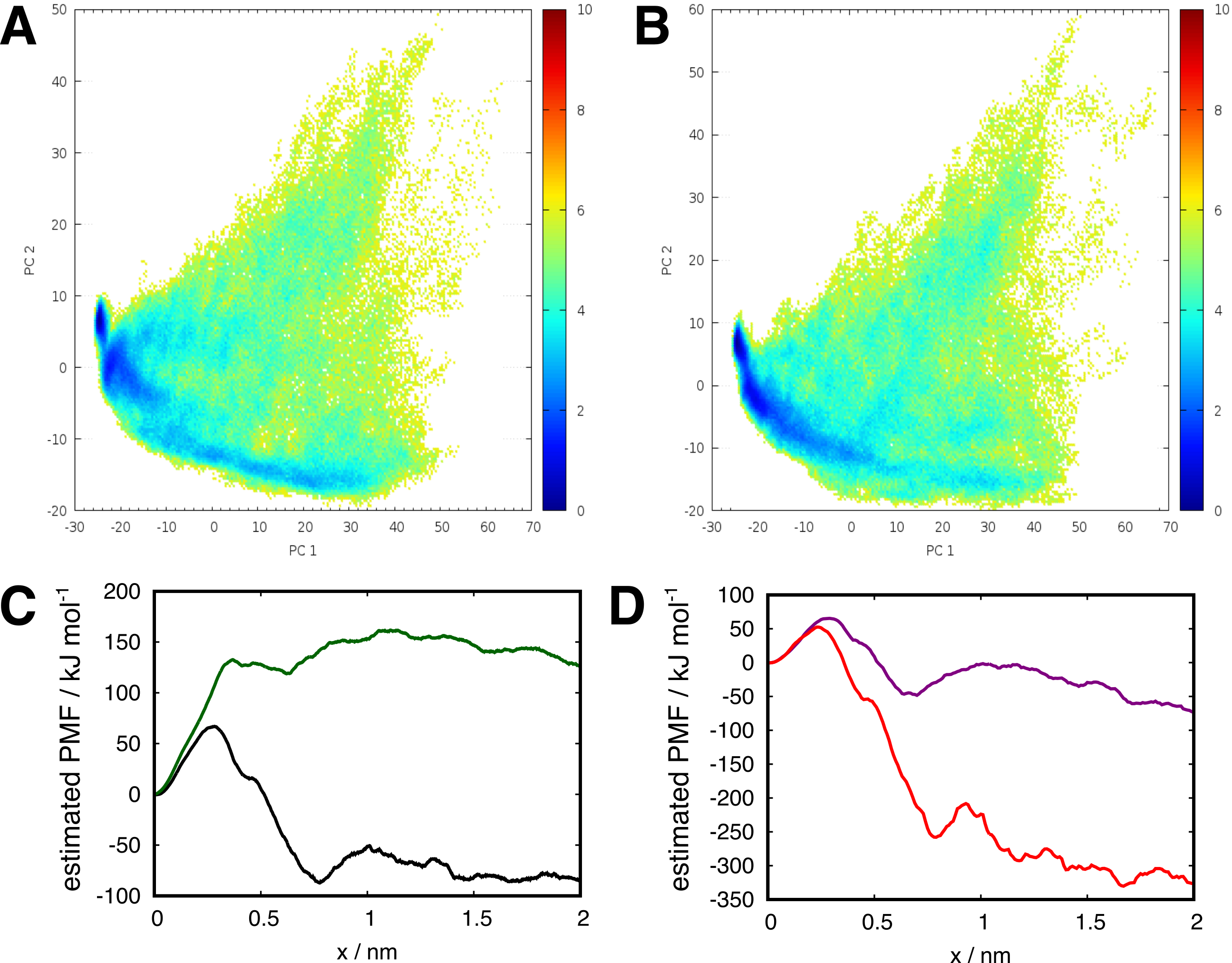}
	\caption{A,B: Biased energy profiles $\Delta \mathcal{G}$ along first two PCs for {\bf 3} (A) and {\bf 4} (B). C: free energies of {\bf 3} (green) and {\bf 4} (black) without pathway separation. D: {\bf 4} pathways in PCA-ML.}
	\label{fig:H24H26}
\end{figure}

To highlight problems one may encounter in the pre-choice of a suitable input coordinate set, we now turn to an interesting case where we found that a small chemical difference between two ligands results in the appearance of an unexpected ligand-internal hidden coordinate.
Compounds {\bf 3} and {\bf 4} share a very similar structure (Fig.~\ref{fig:Intro}b), differing only in the replacement of the amide moiety in {\bf 3} with a sulfonamide in {\bf 4}. As can be seen in Fig.~\ref{fig:H24H26}, both ligands appear to exhibit similar biased energy $\Delta \mathcal{G}$ profiles. A dissipation correction as given in Fig.~\ref{fig:H24H26}c however shows that despite the small chemical difference, {\bf 3} does not exhibit the friction overestimation artefact even without applying pathway separation, while {\bf 4} clearly does.
Furthermore, PCA-ML on {\bf 4} fails to remove the artefact, which still is present for both possible pathways displayed in Fig.~\ref{fig:H24H26}, with RMSD clustering failing as well. 

\begin{figure}[htb]
	\centering
	\includegraphics[width=0.75\textwidth]{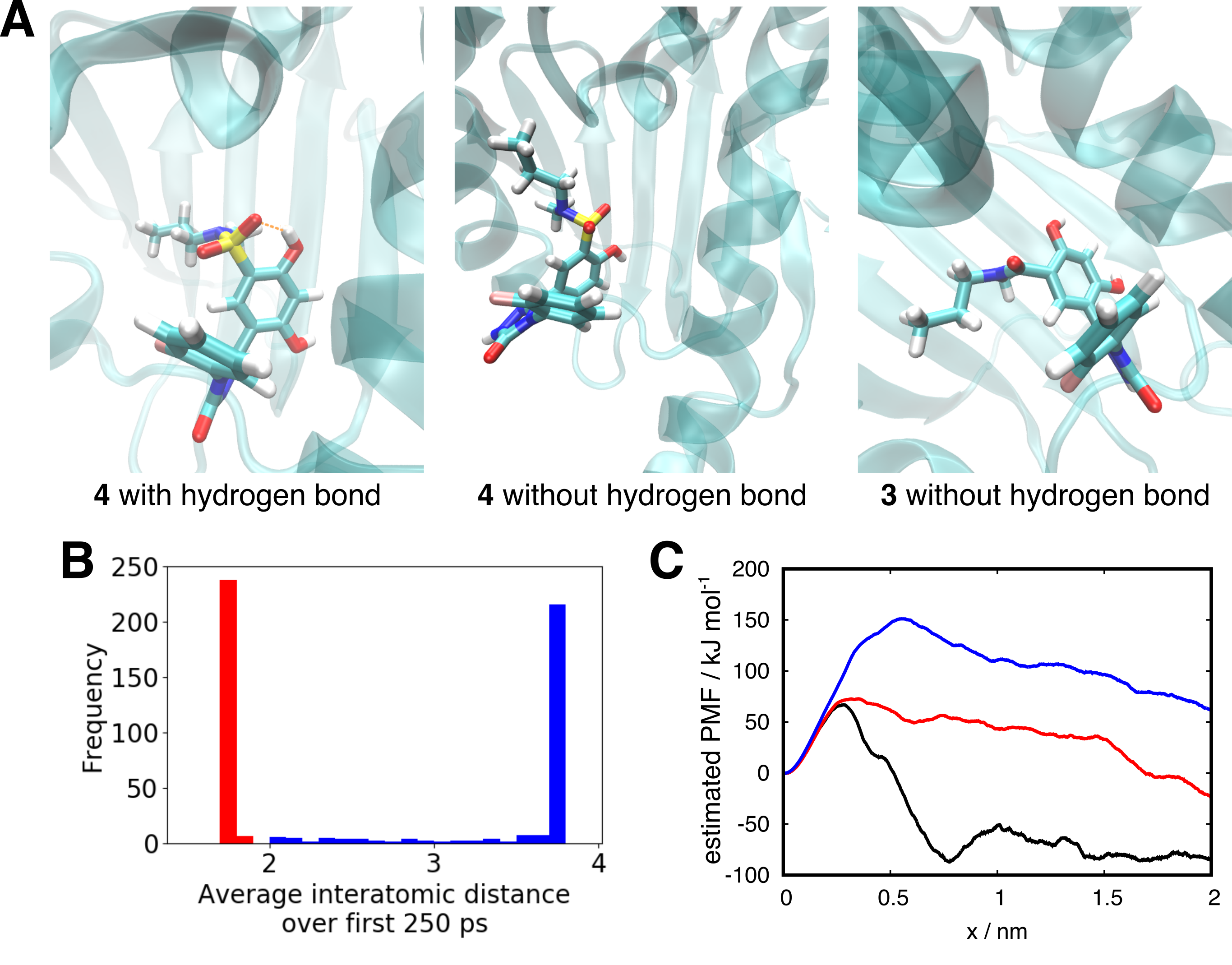}
\caption{A: Structures of ligands {\bf 3} and {\bf 4} during unbinding. B: separation of trajectories with (red) and without (blue) ligand-internal hydrogen bond. C: estimated potential of mean force for compound {\bf 4}, unseparated (black), with (red) and without (blue) hydrogen bond.}
\label{fig:H24H26_res}
\end{figure}

Searching for a suitable hidden coordinate to perform pathway separation, visual inspection of trajectories (Fig.~\ref{fig:H24H26_res}a) revealed a key difference between the two chemically similar ligands: {\bf 4} can form an internal hydrogen bond between the sulfonyl and one of the hydroxyl groups on the adjacent resorcinol ring. When the hydrogen bond is present, rigidity is enforced on the ligand, while a contact with surrounding water can break the hydrogen bond and result in increased conformational dynamics during dissociation. Such a hydrogen bond does not exist with the amide group in {\bf 3} due to the lower electrostatic charge on the carbonyl vs. the sulfonyl oxygen atoms (-0.58~$e$ vs. -0.66~$e$). Thus, we carried out a separation of trajectories into two classes according to formation (or lack thereof) of this internal hydrogen bond as shown in Fig.~\ref{fig:H24H26_res}b and Tab.~\SIH during the first 0.25~ns of simulation, i.e., before the transition barrier in Fig.~\ref{fig:H24H26_res}c. If the average hydrogen-oxygen distance during this time period between the investigated groups was lower than 2 \AA, the hydrogen bond was considered to be present.
As can be seen in the $\Delta G$ curves in Fig.~\ref{fig:H24H26_res}c, the friction overestimation artefact is no longer evident in both of the two resulting pathways in hydrogen bond distance. For the pathway without hydrogen bond, the profile appears very similar to that for {\bf 3}, as would be expected. We note that the $\Delta G$ profile of the pathway with hydrogen bond still exhibits a drop in the potential of mean force in the final 0.5~ns, probably from an additional hidden coordinate that we do not resolve here.

\section{Conclusion}

The aim of this project was to develop and test methods for clustering dcTMD trajectories according to unbinding paths, in order to allow a dissipation correction.
conPCA allowed identification of pathways for trypsin and Hsp90. On this basis, a machine learning model could be built for Hsp90, which automated the classification of trajectories to one of the pathways that were identified by visual inspection. 
Applying this method, dubbed PCA-ML, facilitates pathway classification for dcTMD compared to purely visual inspection.
The other main method applied to characterize the ligand dissociation route was RMSD-based clustering. 
The method is capable of resolving sub-pathways contained within those produced by PCA-ML. 
From this result was reasoned that while PCA-ML is effective at identifying pathways through a protein in general, RMSD clustering gives superior performance when determining the composition of the trajectory classes.
In particular, the ability of neighbor-nets to resolve ambiguity in the input data improves the attribution of trajectories to pathways.
For carrying out a pathway identification, the following workflow is proposed: first of all, as many trajectories as possible should be collected ($\gtrsim$500), and conPCA performed. If it is possible, on this basis, to identify clearly separated pathways, a machine-learning model should be built to classify trajectories. In parallel, RMSD clustering should be performed, and the composition of the classes produced by both classifications compared. If it is not possible to perform PCA-ML, the RMSD pathways should at least be plotted in the PCA space, or the trajectories inspected using graphics software, to ensure that they have a reasonable physical meaning. If PCA-ML and RMSD return pathway classes with similar composition, this is a good sign that the pathway attribution is sound. However, the composition as determined by RMSD should be considered more trustworthy than that produced by PCA-ML, because it takes ligand conformation changes into account. 

Closing this work, we are aware that both PCA-ML and RMSD-based clustering as presented here are tools that can assist a researcher with classifying trajectories into unbinding paths, but require a large amount of human input in the pre-choice of trajectories in training the machine model or in deciding on boundaries in neighbor-nets. The resulting paths are therefore still potentially affected by human bias. In coming works, we will therefore evaluate clustering algorithms that all but eliminate human input, and reduce the number of free parameters to a minimum. Additionally, we will investigate how the potential of mean forces along single pathways can be combined to a global free energy of unbinding. Lastly, a researcher performing a pathway analysis has to be aware that already small chemical differences between ligands can lead to clearly different unbinding pathways, and that the search for the underlying hidden coordinate can be tedious and requires visual inspection as well as considerable chemical intuition. While this insight appears to be frustrating, it relates to the general problem and difficulties of how to identify reaction coordinates for biomolecular processes\cite{Sittel2018a}, which is one of the major current topics in biomolecular simulations method development. Though we still have to find an improved measure for the goodness of a path separation, persistence of the friction overestimation artefact is an indicator that a tested separation is not suitable.

\section{Data and software availability}

dcTMD analysis scripts as well as the fastpca and xgbAnalysis program packages are available at \url{https://www.moldyn.uni-freiburg.de/software.html}. Trypsin-benzamidine start structures, topologies and simulation parameters are contained in a dcTMD tutorial available at \url{https://github.com/floWneffetS/tutorial_dcTMD}. Hsp90 simulation start structures and system topologies are available from the authors upon request.

\begin{acknowledgement}

This work has been supported by the Deutsche
Forschungsgemeinschaft (DFG) via grant WO 2410/2-1 
within the framework of the Research Unit FOR~5099 
''Reducing complexity of nonequilibrium'' (project No.~431945604), 
the High Performance and Cloud
Computing Group at the Zentrum f\"ur Datenverarbeitung of the
University of T\"ubingen, the state of Baden-W\"urttemberg through bwHPC and the DFG through grant no INST 37/935-1 FUGG (RV bw16I016) and the Freiburg Institute for Advanced Studies (FRIAS) of the Albert-Ludwigs-University Freiburg. The authors are grateful to Daniel Nagel for providing additional data evaluation scripts, and to Gerhard Stock, Matthias Post and Moritz Sch\"affler (both University of Freiburg) for helpful discussions.

\end{acknowledgement}

\begin{suppinfo}

One Supplementary Figure (Fig.~\SItrajdistmat) with an example RMSD matrix, one Supplementary Figure with Sankey plots comparing trypsin clusters (Fig.~\SITrypsankey), one Supplementary Figure with thermodynamic integration unbinding free energies of Hsp90 compound {\bf 1} (Fig.~\SItrueG), four Supplementary Figures detailing on the results from conPCA of Hsp90 compound {\bf 1} (Figs.~\SIpcaaas\ to \SIpccomp), two Supplementary Figures (Figs.~\SImlacurr\ and \SImlcontacts) and one Supplementary Table (Tab.~\SItrajdistmat) with information on most important residues in the machine learning procedure, and one Supplementary Table (Tab.~\SIH) with statistics on path separation of compounds {\bf3} and {\bf 4}. (PDF). 

One Supplementary Movie with a 3D isosurface plot of biased energies $\Delta \mathcal{G}$ of compound {\bf 1} in the first three PCs from conPCA with the two contained pathways (MPEG).

\end{suppinfo}



\providecommand{\latin}[1]{#1}
\providecommand*\mcitethebibliography{\thebibliography}
\csname @ifundefined\endcsname{endmcitethebibliography}
  {\let\endmcitethebibliography\endthebibliography}{}


\newpage

\setcounter{figure}{0}   

\renewcommand{\tablename}{Table S}
\renewcommand{\figurename}{Figure S}

\section{Supplementary Figures}

\section{Supplementary Information}

\begin{figure}[H]
\centering
\includegraphics[width=0.4\textwidth]{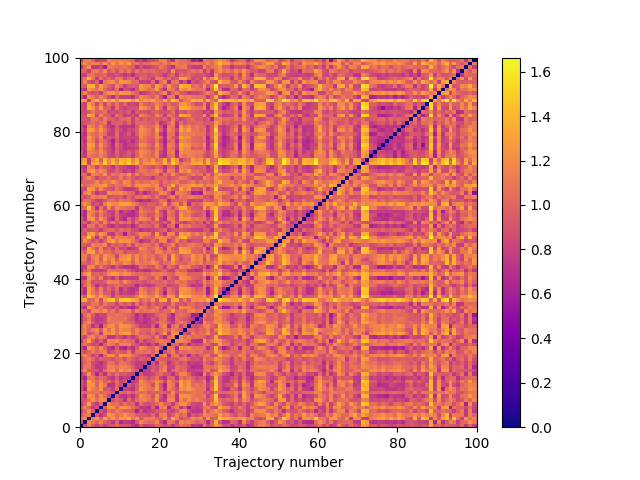}
\caption{Representation of distance matrix of 100 clustered Hsp90 compound {\bf 1} trajectories as a color map, for illustrative purposes. Each pixel represents the mean RMSD distance in nm, averaged over time, between the ligand heavy atoms in two trajectories. Lighter colors indicate greater distances.}
\label{fig:trajdistmat}
\end{figure}

\begin{figure}[H]
\centering
\begin{subfigure}{0.3\textwidth}
\includegraphics[width=\textwidth]{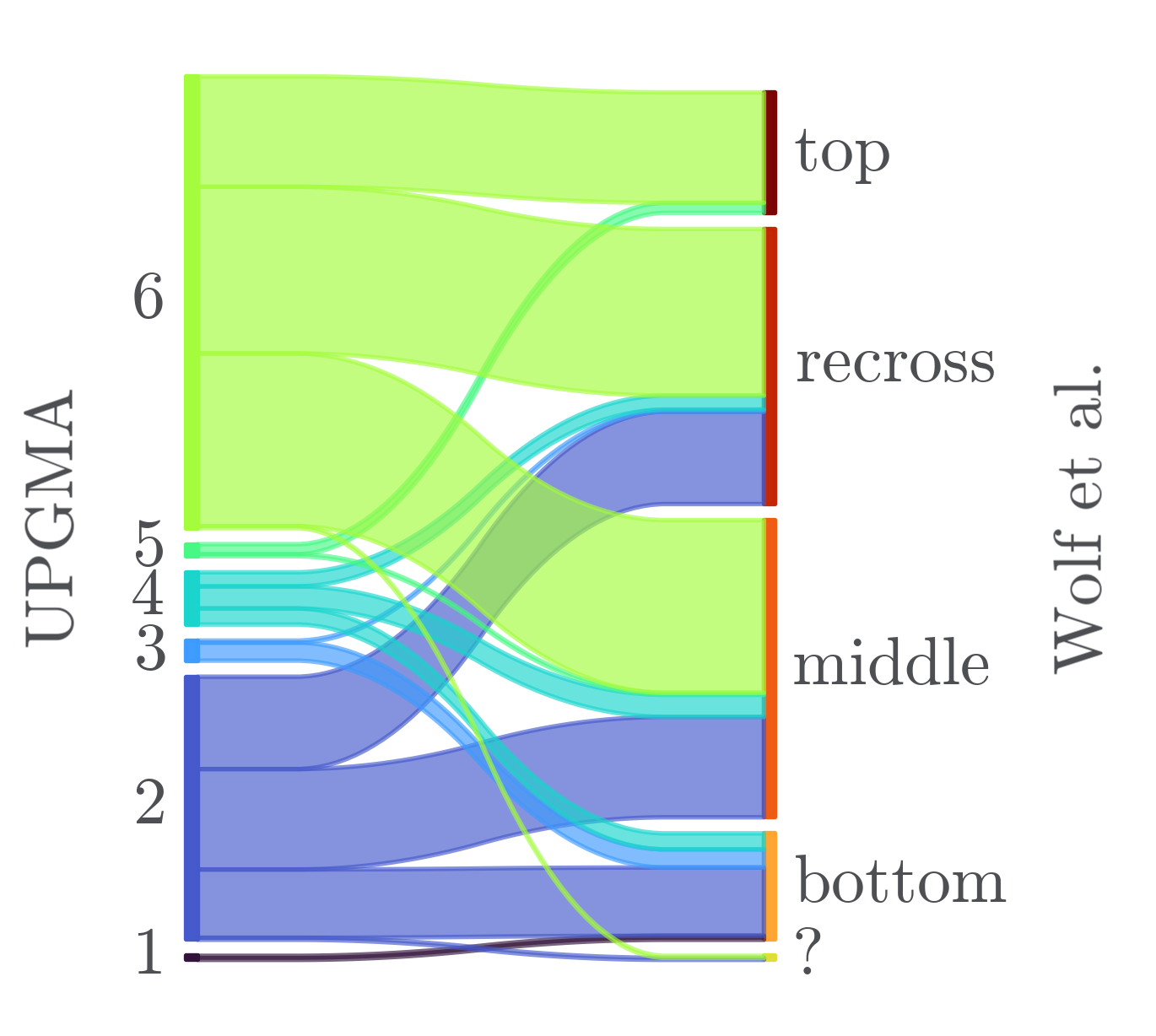}
\caption{}
\end{subfigure}%
\begin{subfigure}{0.35\textwidth}
\centering
\includegraphics[width=\textwidth]{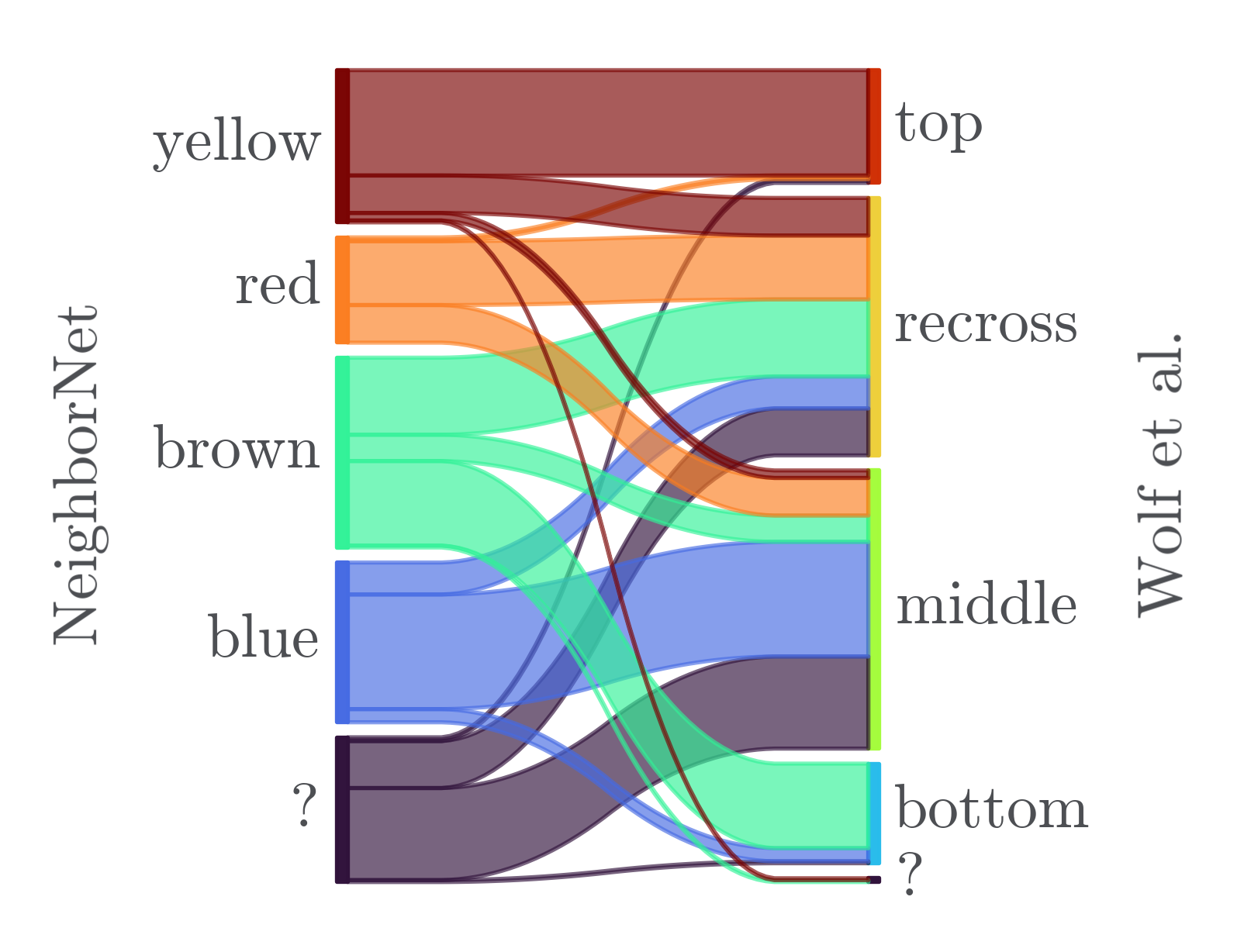}
\caption{}
\end{subfigure}%
\begin{subfigure}{0.3\textwidth}
\centering
\includegraphics[width=\textwidth]{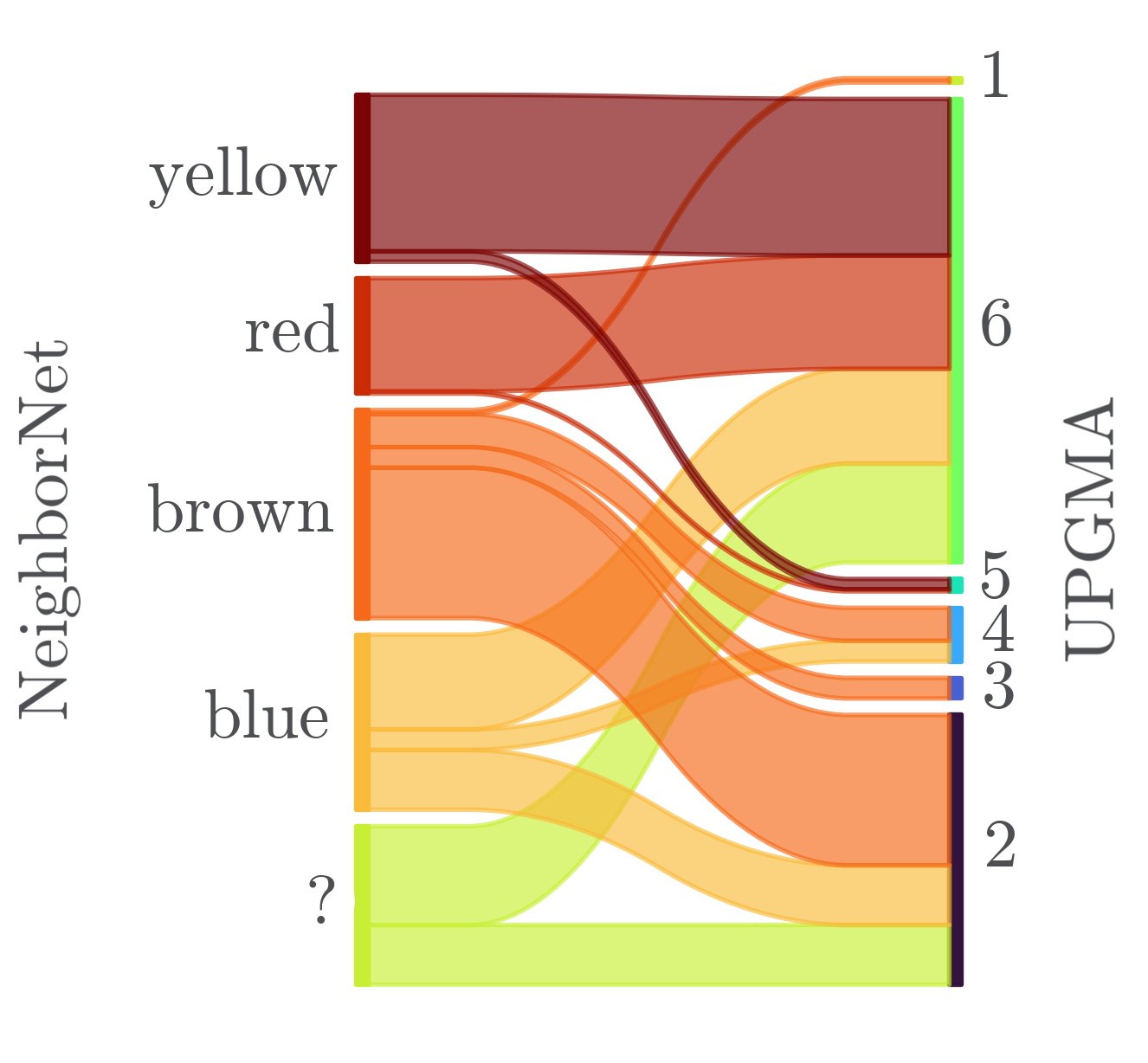}
\caption{}
\end{subfigure}%
\caption{Sankey plots of connection between UPGMA-, neighbor-net- and visual inspection-based (Wolf et al., Nat. Commun. 2020) trajectory clustering in the trypsin-benzamidine complex. a: UPGMA vs. visual inspection. b: Neighbor-net vs. visual inspection. c: Neighbor-net vs. UPGMA.} 
\label{fig:TrypSankey}
\end{figure}

\newpage

\begin{figure}[H]
\centering
\includegraphics[width=0.5\textwidth]{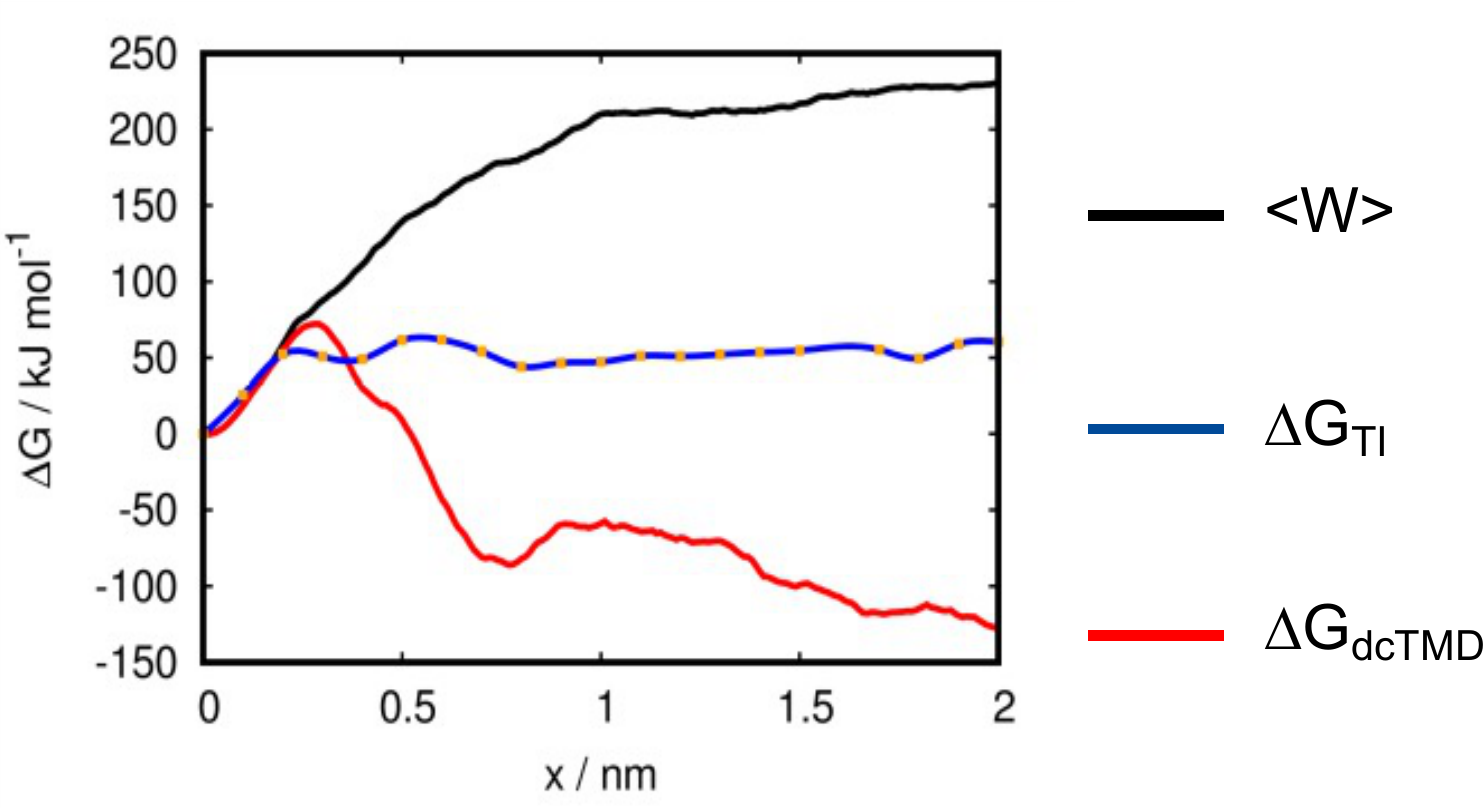}
\caption{Mean work of Hsp90 compound {\bf 1} in black vs. free energy profile from thermodynamic integration (blue) and dcTMD free energy profile for all pulling trajectories affected by the friction overestimation artefact. TI integration window positions in orange. The TI unbinding free energy is on the order of $\sim$50--60~kJ/mol.}
\label{fig:trueG}
\end{figure}

\begin{figure}[H]
\centering
\includegraphics[width=0.3\textwidth]{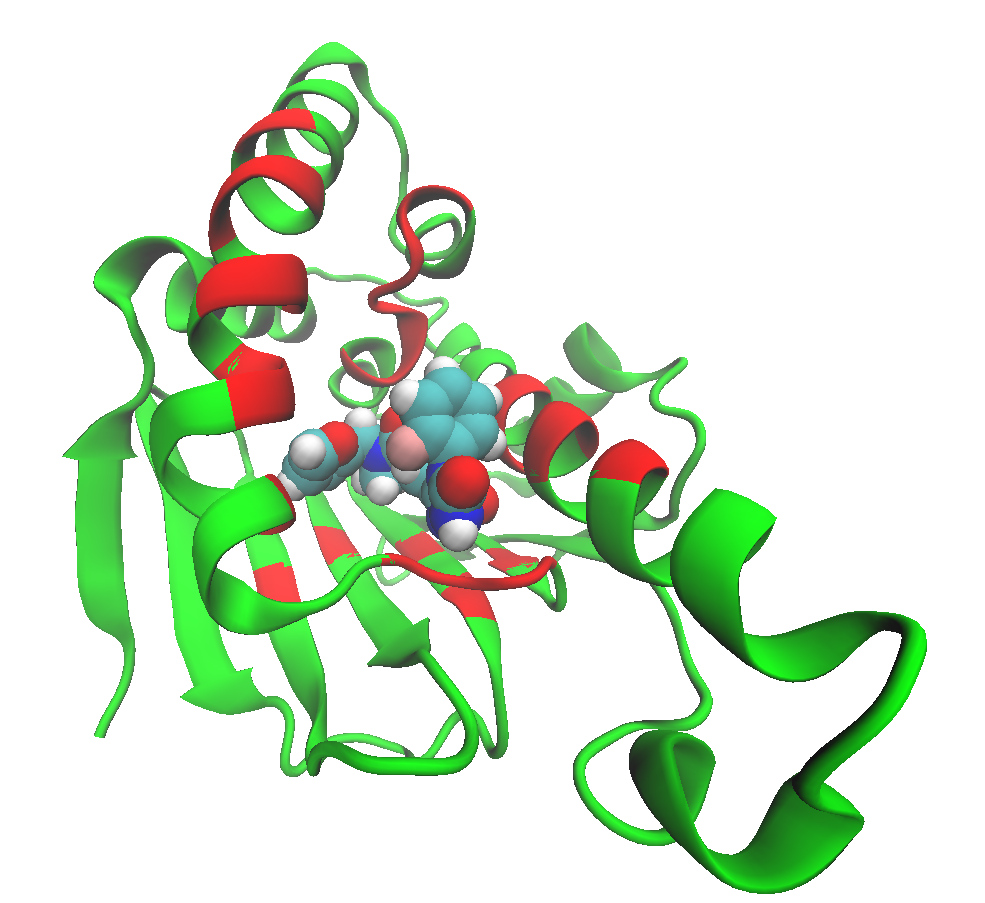}
\caption{Selected contacts for Hsp90 compound {\bf 1} conPCA (red): Asn51, Ser52, Asp54, Ala55, Lys58, Asp93, Ile96, Gly97, Met98, Leu103, Leu107, Gly108, Ile110, Ala111, Ser113, Gly114, Ala117, Phe134, Gly135, Val136, Gly137, Phe138, Tyr139, Val150, Trp162, Thr184, Val186. \label{fig:pca_aas}}
\end{figure}

\begin{figure}[H]
\centering
\begin{subfigure}{0.3\textwidth}
\includegraphics[width=\textwidth]{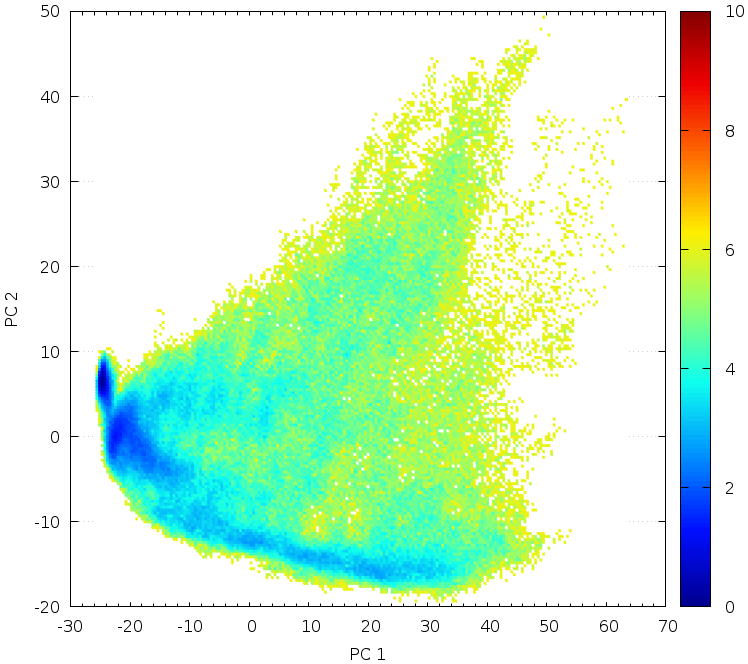}
\caption{}
\end{subfigure}%
\begin{subfigure}{0.3\textwidth}
\centering
\includegraphics[width=\textwidth]{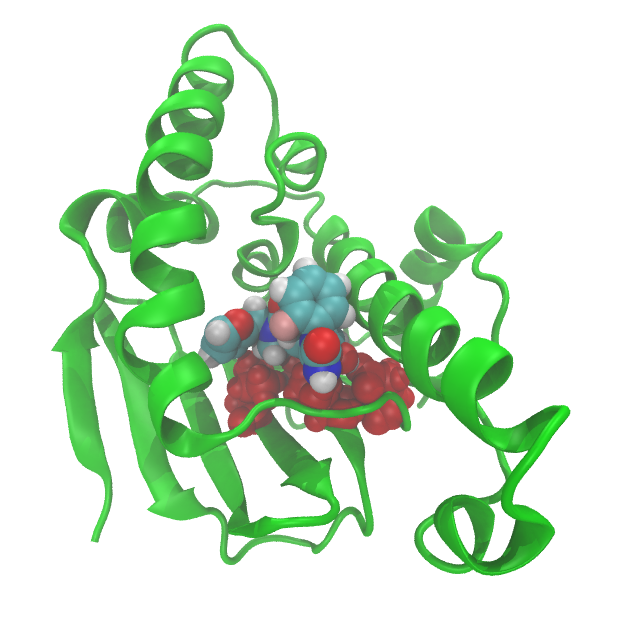}
\caption{}
\end{subfigure}%
\begin{subfigure}{0.3\textwidth}
\centering
\includegraphics[width=\textwidth]{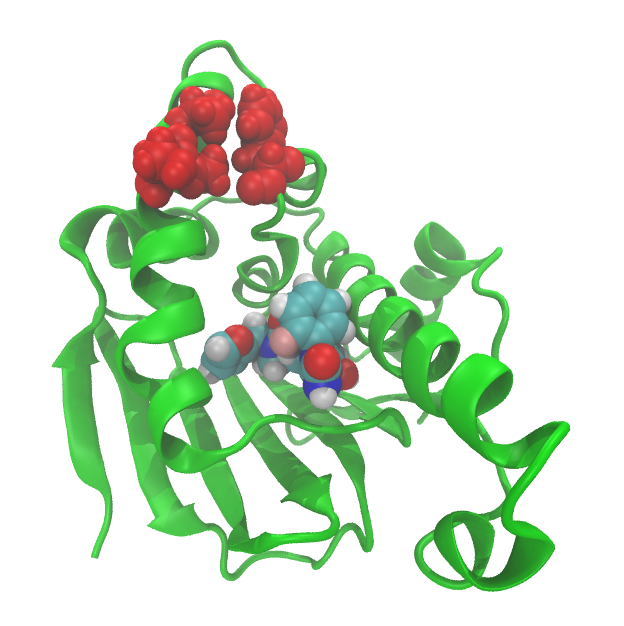}
\caption{}
\end{subfigure}%
\caption{a: Biased energy diagram $\Delta \mathcal{G}$ of the first two PCs for Hsp90 ligand {\bf 1} (100 trajectories). Coloring represents the value of $\Delta \mathcal{G}$. b, c: top residues representing PC1 (a) and PC2 (2) in the respective eigenvector content from Fig.~\SIpccomp.} 
\label{fig:pc_str}
\end{figure}

\newpage

\begin{figure}[htbp]
\centering
\includegraphics[width=0.9\textwidth]{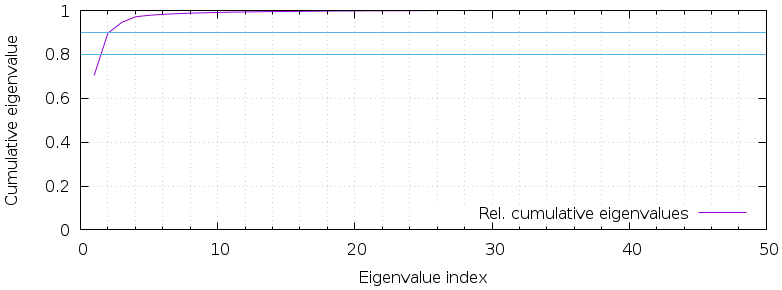}
\caption{Cumulative eigenvalues for Hsp90 ligand {\bf 1} conPCs. \label{fig:eig}}
\end{figure}

\begin{figure}[htbp]
\begin{subfigure}{0.9\textwidth}
\centering
\includegraphics[width=\textwidth]{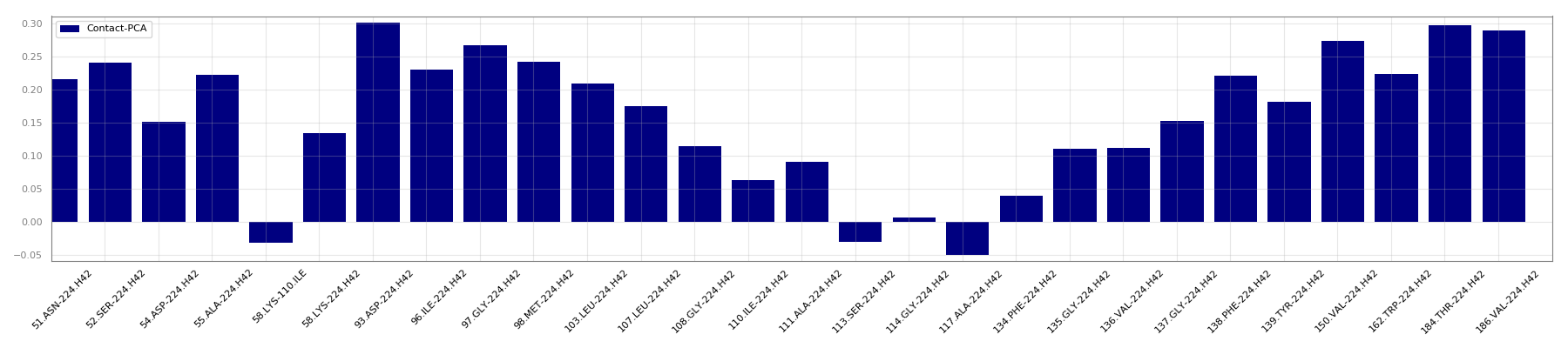}
\caption{PC1}
\end{subfigure} \\
\begin{subfigure}{0.9\textwidth}
\centering
\includegraphics[width=\textwidth]{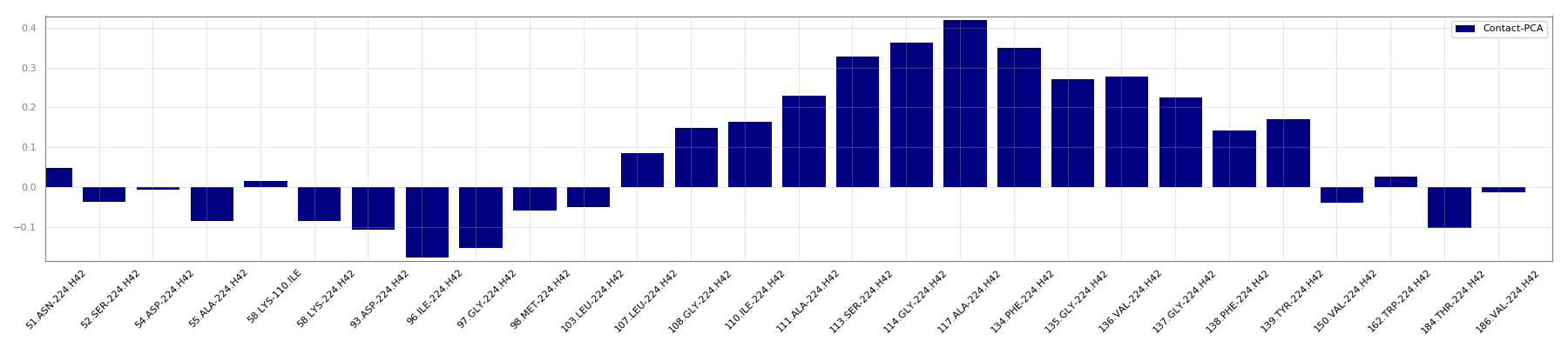}
\caption{PC2}
\end{subfigure}%
\caption{Contribution of each contact to Hsp90 ligand {\bf 1} PCs 1 and 2.}
\label{fig:pc_comp}
\end{figure}

\newpage

\begin{figure}[H]
	\centering
	\includegraphics[width=0.8\textwidth]{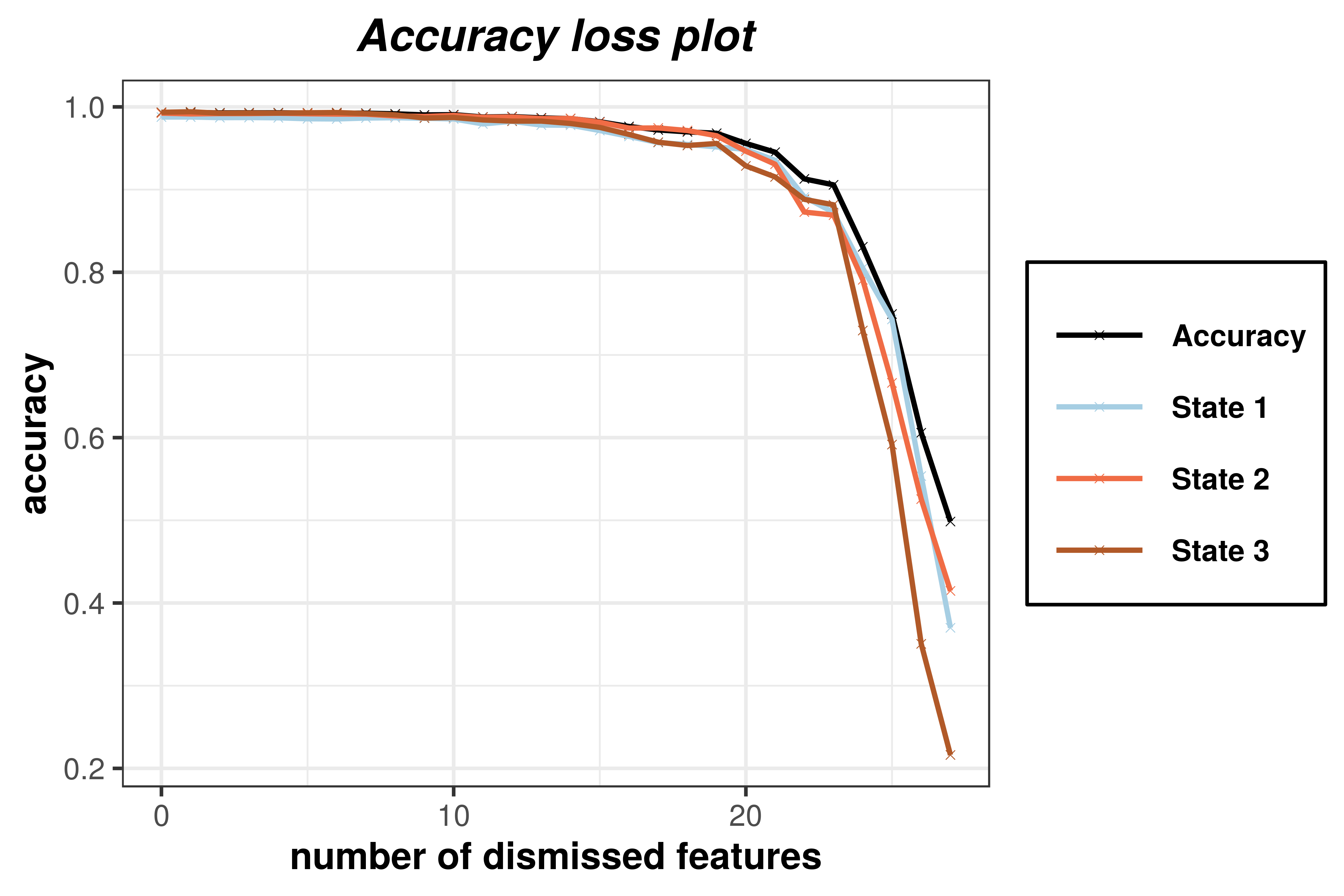}
	\caption{Machine Learning feature sensitivity for path separation of Hsp90 compound {\bf 1}. Features were discarded according to increasing model importance. State 1 and 2 correspond to paths 1 and 2, respectively. State 3 is a neutral class corresponding to PC1 $<-15$ (see Methods).} 
	\label{fig:mlacurr}
\end{figure}

\begin{figure}[H]
	\centering
	\includegraphics[width=0.6\textwidth]{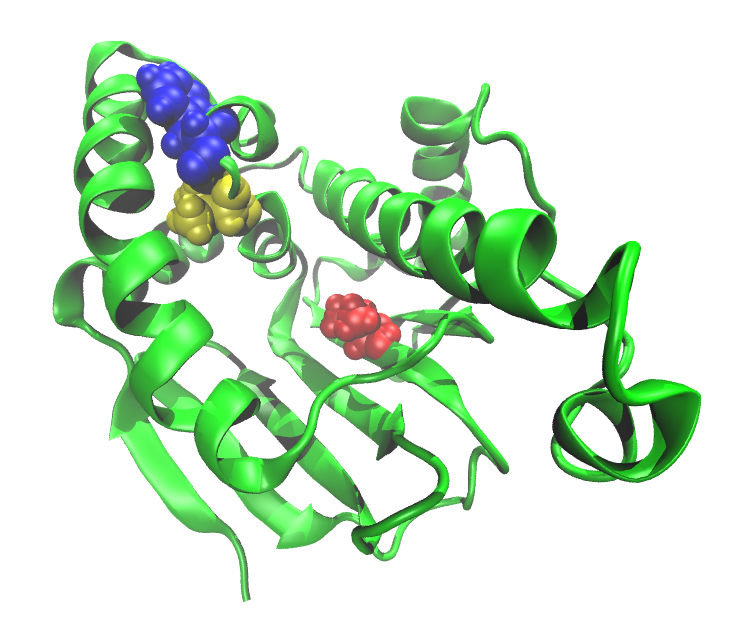}
	\caption{Most important machine learning feature location in Hsp90 compound {\bf 1}. Phe134 (blue), Val136 (yellow) and Val186 (red) in the structure of Hsp90. (For clarity the ligand is not shown.)} 
	\label{fig:mlcontacts}
\end{figure}

\newpage

\section{Supplementary Tables}

\begin{table}[htbp]
\centering
\begin{tabular}{cc}
\hline
Amino acid contact & Importance value \\ \hline
Val186 & 0.31 \\ 
Val136 & 0.29 \\
Phe134 & 0.26 \\
Ala117 & 0.04 \\
Thr184 & 0.02\\ 
\hline
\end{tabular}
\caption{ML importance values for Hsp90 compound {\bf 1} first five contacts.}
\label{table:ml_importance}
\end{table}

\begin{table}[htbp]
	\centering
	\begin{tabular}{cc}
		\hline
		Ligand/path & No. of trajectories \\ \hline
		{\bf 3} & 100   \\
		{\bf 4} (all) & 531  \\
		{\bf 4} (no H-bond) & 284 \\
		{\bf 4} (with H-bond) & 247  \\
		\hline
	\end{tabular}
	\caption{Statistics for Hsp90 compound {\bf 3} and {\bf 4} pathway separation.}
	\label{table:H26}
\end{table}

\end{document}